\documentclass[10pt,aps,pra,twocolumn,showpacs,superscriptaddress,nobalancelastpage,notitlepage,longbibliography]{revtex4-1}

\usepackage{graphicx}
\usepackage{amsmath,amssymb,amsfonts,dsfont}
\usepackage{color}
\usepackage[normalem]{ulem} 
\usepackage{multirow}
\usepackage{array}
\usepackage{upgreek}
\usepackage{mathrsfs}
\usepackage{customcommands}

\begin{document}

\title{Two-qubit spectroscopy of spatiotemporally correlated quantum noise \\ 
in superconducting qubits}

\author{Uwe von L\"upke}
\altaffiliation{Current address: Department of Physics, ETH Z\"urich, 8093 Z\"urich, Switzerland}
\affiliation{Research Laboratory of Electronics, Massachusetts Institute of Technology, Cambridge, MA 02139, USA}

\author{F\'elix Beaudoin}
\affiliation{Department of Physics and Astronomy, Dartmouth College, Hanover, NH 03755, USA}
\affiliation{Nanoacademic Technologies Inc., 666 rue Sherbrooke Ouest, Suite 802, Montr\'eal, Qu\'ebec, Canada H3A 1E7}

\author{Leigh M. Norris}
\affiliation{Department of Physics and Astronomy, Dartmouth College, Hanover, NH 03755, USA}

\author{Youngkyu Sung}
\affiliation{Research Laboratory of Electronics, Massachusetts Institute of Technology, Cambridge, MA 02139, USA}

\author{Roni Winik}
\affiliation{Research Laboratory of Electronics, Massachusetts Institute of Technology, Cambridge, MA 02139, USA}

\author{Jack Y. Qiu}
\affiliation{Research Laboratory of Electronics, Massachusetts Institute of Technology, Cambridge, MA 02139, USA}

\author{Morten Kjaergaard}
\affiliation{Research Laboratory of Electronics, Massachusetts Institute of Technology, Cambridge, MA 02139, USA}

\author{David Kim}
\affiliation{MIT Lincoln Laboratory, 244 Wood Street, Lexington, MA 02421, USA}

\author{Jonilyn Yoder}
\affiliation{MIT Lincoln Laboratory, 244 Wood Street, Lexington, MA 02421, USA}

\author{Simon Gustavsson}
\affiliation{Research Laboratory of Electronics, Massachusetts Institute of Technology, Cambridge, MA 02139, USA}

\author{Lorenza Viola}
\affiliation{Department of Physics and Astronomy, Dartmouth College, Hanover, NH 03755, USA}

\author{William D. Oliver}
\affiliation{Research Laboratory of Electronics, Massachusetts Institute of Technology, Cambridge, MA 02139, USA}
\affiliation{MIT Lincoln Laboratory, 244 Wood Street, Lexington, MA 02421, USA}
\affiliation{Department of Physics, Massachusetts Institute of Technology, Cambridge, MA 02139, USA}

\date{\today}

\begin{abstract}
Noise that exhibits significant temporal and spatial correlations across multiple qubits can be especially harmful to both fault-tolerant quantum computation and quantum-enhanced metrology. However, a complete spectral characterization of the noise environment of even a two-qubit system has not been reported thus far. We propose and experimentally validate a protocol for two-qubit dephasing noise spectroscopy based on continuous control modulation. By combining ideas from spin-locking relaxometry with a statistically motivated robust estimation approach, our protocol allows for the simultaneous reconstruction of all the single-qubit and two-qubit cross-correlation spectra, including access to their distinctive non-classical features. Only single-qubit control manipulations and state-tomography measurements are employed, with no need for entangled-state preparation or readout of two-qubit observables. While our experimental validation uses two superconducting qubits coupled to a shared engineered noise source, our methodology is portable to a variety of dephasing-dominated qubit architectures. By pushing quantum noise spectroscopy beyond the single-qubit setting, our work paves the way to characterizing spatiotemporal correlations in both engineered and naturally occurring noise environments. 
\end{abstract}

\maketitle

\section{Introduction}

Quantum information science is poised to deliver unprecedented opportunities in terms of both fundamental physics and device technologies, by pushing existing boundaries in areas as diverse as quantum computation and simulation, secure communication, and quantum-enhanced sensing and metrology. Notably, advances in quantum control and systems engineering are enabling access to intermediate-scale quantum processors whose capabilities are beyond what may be tractable classically \cite{PreskillNISQ}, with impressive achievements having recently been reported \cite{Lukin2017,Monroe2017,IonQ,Google}. Ultimately, however, realizing the full potential of these technologies will crucially depend on sustained progress in characterizing and overcoming the effects of noise that limit qubit performance. 

In the context of entanglement-assisted quantum metrology, spatial and temporal correlations of the noise dictate the extent by which the standard quantum limit on precision may be overcome in the estimation of physical parameters~\cite{HaaseReview,SmerziReview}, in particular when the quantum-mechanical nature of the environment must be explicitly accounted for~\cite{beaudoin2018ramsey}. Once characterized, spatial noise correlations may be exploited for augmenting the performance of quantum sensors via tailored quantum encoding \cite{dorner2012quantum} or error correction \cite{Layden_npj}. Likewise, quantitative knowledge about noise properties and their correlations may prove instrumental in designing optimized quantum-control or error-mitigation strategies that can be substantially more efficient than general-purpose schemes \cite{multiDD,Layden_qec}, as well as in determining optimal parameter regimes for ``spectator qubits'' to be useful in improving control performance of proximal data qubits \cite{BrownSpectator}. 

Ultimately, noise correlations will play a key role in determining the feasibility of large-scale fault-tolerant quantum computation: establishing that noise correlations decay sufficiently rapidly in space and time is central for validating the locality assumptions under which the existence of an accuracy threshold may be rigorously derived beyond the paradigm of independent errors \cite{ng2009fault,PreskillSufficient}. In turn, the structure of the noise, including the relevance of correlated error processes, is expected to strongly influence the value of the error threshold itself and inform ways in which resource-optimized architectures may be designed \cite{Flammia2018,Brown2019}. As a result, developing viable methodologies to detect and simultaneously characterize both spatial and temporal correlations present in realistic multiqubit noise environments is an imperative next step.

In a single-qubit setting, temporal correlations of dephasing noise that may be assumed to be stationary and Gaussian are characterized in the frequency domain by a single noise spectrum -- namely, the Fourier transform of the two-point correlation function of the noise operator with respect to the time lag \cite{clerk2010introduction}. Estimation of the spectrum from experimental data may be achieved through various ``quantum noise spectroscopy'' protocols, which employ either pulsed or continuous control modulation of the qubit sensor to suitably shape its spectral response. To date, measurements of the noise spectrum have been reported across a wide variety of experimental qubit platforms, including nuclear magnetic resonance~\cite{alvarez2011measuring,willick2018efficient}, superconducting quantum circuits~\cite{bylander2011noise,yuge2011measurement,yan2013rotating,quintana2017observation,yan2018distinguishing}, nitrogen-vacancy centers~\cite{bar2012suppression,romach2015}, spin donors in semiconductors~\cite{chan2018assessment}, and trapped ions~\cite{frey2017application,frey2019}.

In a multi-qubit setting, complete characterization of Gaussian dephasing noise necessitates estimation of the full set of spectra $\{S_{jk}(\w)\}$, defined by the Fourier transform of the correlation functions of noise operators acting on each possible combination of qubits $j$ and $k$. While temporal noise correlations that affect qubits individually are now described in terms of \emph{self-spectra} $S_{jj}(\w)$, coexisting spatial and temporal correlations are captured by the two-qubit \emph{cross-spectra} $\{S_{jk}(\w)\}$, with $j\neq k$. Spatial noise correlations have been probed and their strength upper-bounded in recent experiments using superconducting fluxonium qubits~\cite{kou2018simultaneous}, nitrogen-vacancy centers in diamond~\cite{bradley2019ten} and spin qubits in semiconductors~\cite{boter2019spatial}. However, all the protocols implemented thus far lack the frequency sensitivity needed for full-fledged multi-qubit spectroscopy of noise that may be in general \emph{spatiotemporally correlated and non-classical}. Despite promising theoretical proposals~\cite{paz2017multiqubit,szankowski2016spectroscopy,krzywda2018dynamical,rivas2015quantifying} as well as an experimental approach using a specific correlation measure \cite{postler2018experimental}, measurements of a two-qubit cross-spectrum remain yet to be reported. 

\begin{figure}[t]
\includegraphics[width=7.5cm]{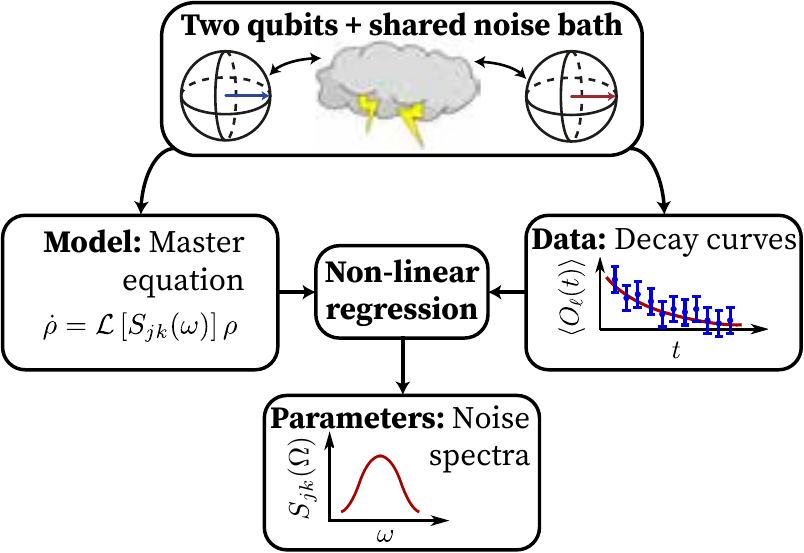}
 \caption{{\em Concept of two-qubit quantum noise spectroscopy experiment}. Two qubits are coupled to a common bath, leading to spatiotemporally correlated noise. The reduced dynamics of the qubit system is modeled through a master equation which  contains all the noise spectra. Using spin-locking control sequences, we measure decay curves of different observables. Comparison between the evolution of these observables predicted by the master equation against measured data yields the target spectra as fit parameters. While our methodology is device-independent, we use superconducting qubits for experimental validation.
\label{figSketch}}
\end{figure}

In this paper, we theoretically develop a quantum-control protocol for two-qubit spectral estimation, and validate it experimentally in a circuit quantum electrodynamics (cQED) system using engineered photon shot noise. Building on continuous-control modulation spin-locking techniques previously implemented in a single-qubit setting \cite{yan2013rotating,yan2018distinguishing}, we formulate the problem in the framework of robust estimation theory~\cite{huber1981robust} and demonstrate the simultaneous reconstruction of all the two-qubit self-spectra and cross-spectra that characterize our engineered noise source within a Gaussian approximation. In contrast to existing proposals employing dynamical-decoupling comb-based noise spectroscopy~\cite{paz2017multiqubit,szankowski2016spectroscopy,krzywda2018dynamical}, our approach does not require the design and application of long sequences of nearly-instantaneous pulses. As an additional advantage, for instance with respect to methods exploiting  decoherence-free subspaces~\cite{boter2019spatial}, no entangled states nor two-qubit gates are needed. Instead, our protocol relies on continuous driving of the individual qubits followed by simultaneous single-qubit readout. We thus anticipate this protocol to be compatible with any multiqubit architecture in which these resources are available.

Figure~\ref{figSketch} illustrates the methodology underlying our two-qubit spectroscopy approach. To appreciate the underlying physical principles, it is useful to contrast our method with single-qubit noise spectroscopy via spin-locking relaxometry~\cite{yan2013rotating,yan2018distinguishing}: there, a microwave tone is applied on the sensor qubit to effectively create a ``dressed qubit,'' whose level splitting equals the frequency of the Rabi oscillations induced by the drive. Since this dressed qubit is predominantly sensitive to the noise spectrum at its transition frequency, the \mbox{(self-)} spectrum may then be sampled, point by point, by measuring the decoherence rate the dressed qubit experiences as a function of the applied Rabi frequency. In our generalized two-qubit spin-locking protocol, we simultaneously drive two sensor qubits with amplitudes set to give an identical Rabi frequency, under ideal conditions. This effectively produces two dressed qubits, which are now sensing the same frequency component of the noise through the self- and cross-spectra that enter the master equation governing their dynamics. These spectra are then reconstructed at each Rabi frequency of interest by fitting the numerical solution of the master equation to experimentally measured decay curves.

The paper is organized as follows. In Sec.~\ref{secMethodology}, we provide the theoretical foundation of the reconstruction protocol we qualitatively described above. While Sec.~\ref{secSpinLocking} presents the derivation of a master equation (ME) for the two-qubit reduced dynamics, Sec.~\ref{secEstimation} is devoted to the robust statistical procedure we employ to extract the spectra from fits of time-dependent expectation values computed from the ME to experimental data. In order to set the stage for the experimental validation of our protocol, we devote Sec.~\ref{secDemo} to the demonstration of engineered correlated photon shot noise in the cQED test-bed we employ (see Fig.~\ref{figMicrograph}), and to the experimental determination of the parameters entering the shot-noise spectra. We first perform a parametric analysis of this correlated noise in our experimental setup (Sec.~\ref{secRamsey}), and then demonstrate our capability to selectively probe a specific frequency region of the cross-spectra (Sec.~\ref{secSpinlockingExperiment}). After discussing in Sec.~\ref{secNonIdealities} some technical modifications needed to adapt our protocol to the non-idealities specific to our circuit-QED platform, our main results are reported in Sec.~\ref{secEstimationResults}: namely, the non-parametric experimental reconstruction of the two-qubit self- and cross-spectra that characterize our engineered noise source. Full technical detail about the derivation of the MEs used to model the reduced dynamics in both the ideal and experimentally relevant parameter regimes is included in Appendix \ref{secTCL}, whereas Appendix \ref{secConfidenceIntervals} and Appendix \ref{secHuber} discuss various aspects pertinent to the robust statistical estimation procedure we employ.

\section{Methodology for two-qubit noise spectroscopy 
\label{secMethodology}}

\subsection{Noise model and reduced master equation for spin-locking dynamics} 
\label{secSpinLocking}

We summarize here the basic ideas of spin-locking relaxometry~\cite{yan2013rotating,yan2018distinguishing} and generalize them to the two-qubit setting. To describe the driven evolution of two qubits under a noisy environment, we consider the Hamiltonian
\begin{align}
 H(t)=H_\mrm{S}(t)+H_\mrm{SB}+H_\mrm B,	
 \label{eqnH}
\end{align}
where $H_\mrm{S}(t)$ is the time-dependent Hamiltonian of the two-qubit system S, $H_\mrm B$ the Hamiltonian of the bath B, and $H_\mrm{SB}$ describes system-bath coupling. We assume the two qubits to be characterized by angular-frequency splittings $\w_{\mrm q j}$, and each to be coherently driven at frequency $\w_{\mrm d j}$ with a drive strength $\W_j$, $j\;\in\;\{1,2\}$. Setting $\hbar\equiv 1$, we thus consider the system Hamiltonian
\begin{align}
 H_\mrm{S}(t)=\sum_{j\in\{1,2\}}\left[\frac{\w_{\mrm qj}}2\s^z_j+\W_j\cos(\w_{\mrm d j}t)\s^x_j\right],	
 \label{eqnHSd}
\end{align}
where $\s^x_j$ and $\s^z_j$ are the Pauli matrices for qubit $j$. Throughout this paper, the $+1$ and $-1$ eigenstates of $\sigma^z_j$ will be denoted by $\ket 1_j$ and $\ket 0_j$, respectively. While we leave $H_\mrm B$ unspecified, we specialize to single-axis system-bath couplings of the form
\begin{align}
 H_\mrm{SB}=\sum_j B_j \s^z_j,	\label{eqnHSB}
\end{align}
where $B_j$ is the bath operator that couples to qubit $j$. 

In the absence of coherent drives ($\W_j=0$, $j=1,2$), $H_\mrm{SB}$ generates pure-dephasing evolution. However, in an appropriate rotating frame, the coherent drive ``tilts'' the quantization axis of each qubit by a $\pi/2$ angle, thus turning dephasing noise into a source of energy absorption and emission at rates that probe the two-qubit spectra at frequencies $\W_j$. To describe this phenomenon and exploit it for noise-spectroscopy applications, we apply the unitary tranformation $R(t)=\exp[-i\sum_j\w_{\mrm dj}t\s^z_j/2]$ to move to a reference frame that rotates at the drive frequencies, and in which the Hamiltonian is $H_\mrm R(t)\equiv R^\dag(t) H(t)R(t)-i R(t)^\dag\dot R(t).$ By effecting the frame transformation and invoking the rotating-wave approximation (RWA) to drop counter-rotating terms oscillating at frequency $2\omega_{\mrm d j}$, the rotating-frame Hamiltonian is then
\begin{align}
H_\mrm R &\approx H'_\mrm{S}+H_\mrm{SB}+H_\mrm B,	\label{eqnHR}
\end{align}
where $H'_\mrm S \equiv \sum_j(\D_{\mrm q j}\s^z_j+\W_j\s^x_j)/2$ and $\D_{\mrm q j}\equiv \w_{\mrm q j}-\w_{\mrm d j}$ is the detuning of drive $j$ from $\w_{\mrm qj}$.

The time-independent Hamiltonian $H'_\mrm S$ can be diagonalized by transforming to a  rotated \emph{spin-locking basis}, spanned by $\ket{{+x}}_j\equiv\cos(\vartheta_j/2)\ket1_j+\sin(\vartheta_j/2)\ket0_j$ and $\ket{{-x}}_j\equiv-\sin(\vartheta_j/2)\ket1_j+\cos(\vartheta_j/2)\ket0_j$, where $\vartheta_j\equiv \mrm{arctan}(\W_j/\D_{\mrm q j})$ is the angle by which the qubit quantization axis is rotated under the drives. In this basis, \begin{align}
 H'_\mrm S &= \sum_j\frac12\sqrt{\D_{\mrm q j}^2+\W_j^2}\;\tau^z_j,	
 \label{eqnHSprime}\\
 H_\mrm{SB} &= \sum_jB_j\left(\cos\vartheta_j\,\tau^z_j-\sin\vartheta_j\, \tau^x_j\right),	
\label{eqnHSBprime}
\end{align}
where we have introduced the Pauli matrices $\tau^x_j\equiv{\ket{{+x}}}\bra{-x}_j+\ket{-x}\bra{+x}_j$ and $\tau^z_j\equiv\proj{{+x}}_j-\proj{{-x}}_j$.  Setting $\D_{\mrm q j}=0\;\forall\;j$ leads to $\vartheta_j=\pi/2\;\forall\;j$, which further simplifies the above Hamiltonians to 
\begin{align}
 H'_\mrm S=\sum_j\frac{\W_j}2\tau^z_j,	\hspace{3mm}H_\mrm{SB}=-\sum_jB_j\tau^x_j.	\label{eqnHSBspinLocking}
\end{align}
In this spin-locking basis, the coherent drives then define two effective ``dressed'' qubits quantized along $\tau^z_j$, with angular-frequency splittings equal to the Rabi frequencies $\W_j$, and subject to purely transverse noise along $\tau^x_j$. 

We may describe the evolution of the dressed qubits in the spin-locking basis by tracing out the bath and deriving a reduced ME for the density operator $\rho(t)\equiv\Tr_\mrm{B}[\rho_\mrm{tot}(t)]$. As detailed in Appendix~\ref{secTCLideal}, to do so we employ a standard time-convolutionless (TCL) ME approach~\cite{breuer2002theory}. We assume an initially separable initial state $\rho_\mrm{tot}(0)\equiv \rho(0)\otimes\rho_\mrm B$, where $\rho(0)$ and $\rho_\mrm B$ are the initial density operator of S and B, respectively. In addition, we consider stationary noise with zero mean, so that $\mean{B_j(t)}=0$ and $\mean{B_j(t)B_k(s)}=\mean{B_j(t-s)B_k(0)}\equiv\mean{B_j(\tau)B_k(0)}\;\forall\;t,s,j,k$, where $B_j(t)\equiv\eul{i H_\mrm B t}B_j\eul{-iH_\mrm B t}$ is the time-dependent noise operator for qubit $j$ in the interaction picture associated to the free bath Hamiltonian, the time lag $\tau\equiv t-s$, and $\langle \cdot\rangle $ denotes expectation with respect to the initial bath state $\rho_\mrm B$. We also assume that the coupling between the system and the bath is weak enough to truncate the TCL generator at second order, and employ a secular approximation to drop terms oscillating with frequency $\W_1+\W_2$. Setting $\W_1=\W_2\equiv \W$, the two dressed qubits are most sensitive to the noise spectra in a frequency window of width $\sim 1/t$ around their splitting $\pm\W$. This enables us to simplify the reduced ME in the limit of a sufficiently long evolution time. For spectra $S_{jk}(\w)$ that vary sufficiently slowly with frequency about $\w=\pm\W$, we consider only the contribution of the spectra at $\omega=\pm \W$, and finally arrive at
\begin{align}
\dot\rho(t)=&-i[H'_\mrm S,\rho(t)]+\sum_{jk}\mathcal L_{jk}\rho(t),	
\label{eqnReduced}
\end{align}
where the superoperators $\mathcal L_{jk}$ are defined by
\begin{align}
  &\mathcal L_{jk}\rho\equiv S_{jk}(-\W)\left[\tau^-_k \rho \tau^+_j-\frac12\left\{\tau^+_j\tau^-_k,\rho\right\}\right]\notag\\
  &\qquad+S_{jk}(\W)\left[\tau^+_k \rho\tau^-_j-\frac12\left\{\tau^-_j\tau^+_k,\rho\right\}\right],	\label{eqnCorrelatedDecay}
\end{align}
with $\{A,B\}\equiv AB-BA$ denoting the anticommutator of $A$ and $B$. The superoperators introduced in Eq.~\eq{eqnCorrelatedDecay} describe \emph{correlated decay and absorption processes} with strength proportional to the two-qubit spectra evaluated at $\w=\pm\W$. These two-qubit spectra are given by
\begin{align}
 S_{jk}(\w)\equiv\int_{-\infty}^{\infty}d\tau\,\eul{-i\w \tau}\mean{B_j(\tau)B_k(0)}.	\label{eqnSpectra}
\end{align}

The reduced ME defined by Eqs.~\eq{eqnReduced}-\eq{eqnCorrelatedDecay} involves \emph{all} the spectra that are needed to characterize stationary noise that may in general be non-classical (in the sense that the commutator $[B_j(t),B_k(s)]\ne 0,$ $t \ne s$), and display 
arbitrary temporal and spatial correlations in a two-qubit system -- provided that noise can be assumed to be Gaussian and acting only along the $\s^z_j$ axes (purely dephasing) in the laboratory frame. The Gaussian assumption may be satisfied exactly, for example for bosonic baths at equilibrium~\cite{leggett1987dynamics}, or in an approximate sense when either a large number of independent bath degrees of freedom are involved~\cite{paladino2014noise} or the coupling is sufficiently weak for any higher-order cumulants to be negligible. It is worth noting that, from a rigorous spectral estimation standpoint, the assumptions of single-axis noise and Gaussianity cannot be expected to be valid \emph{a priori}, and should always be verified experimentally through spectroscopic means~\cite{norris2016qubit,sung2019non,paz-silva2019extending}.

\subsection{Robust estimation of two-qubit spectra} 
\label{secEstimation}

Since Eqs.~\eq{eqnReduced}-\eq{eqnCorrelatedDecay} contain all the spectra of interest for the present two-qubit problem, an appropriate choice of experimental observations can enable us to infer $S_{jk}(\W)$ for arbitrary $j,k\,\in\,\{1,2\}$. For notational convenience, for a given Rabi frequency $\W$, we collect the two-qubit spectra that we aim to estimate into a \emph{spectrum vector}, 
\begin{eqnarray*}
&& \mvec S(\W)\equiv\{S_{11}(\W), S_{22}(\W), \mrm{Re}[S_{12}(\W)],\mrm{Im}[S_{12}(\W)], \\
&& \quad S_{11}(-\W), S_{22}(-\W),\mrm{Re}[S_{12}(-\W)],\mrm{Im}[S_{12}(-\W)]\}^T.
\end{eqnarray*} 
To devise a protocol for estimation of $\mvec S$, we will adopt an inverse-problems perspective and infer $\mvec S$ by performing a non-linear regression that fits numerical solutions of the reduced ME to experimental data.

\begin{figure}[thbp]
\begin{center}
\includegraphics[width=0.99\columnwidth]{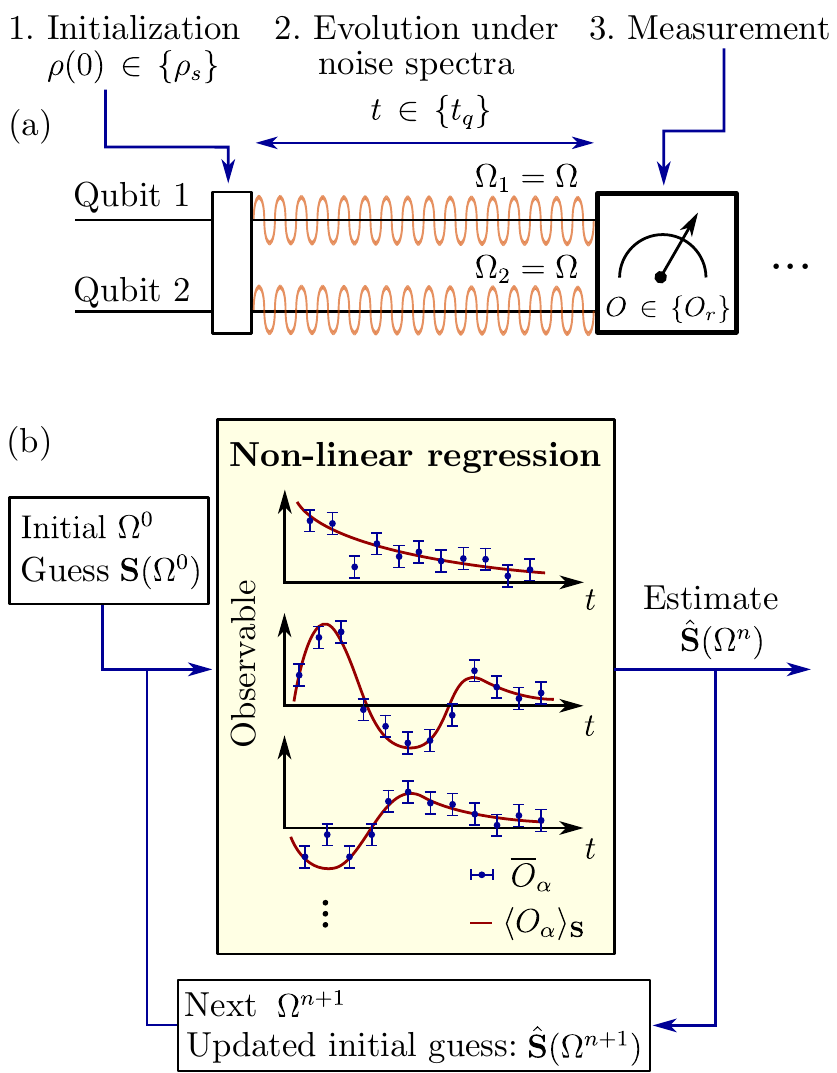}
\end{center}
\caption{{\em Protocol for spectroscopy of spatiotemporally correlated noise}. {\bf (a)} Control and measurement cycle. The qubit system S is initialized in state $\rho(0)\,\in\,\{\rho_s\}$,  where $s\in\{1,2,\ldots,N_\mrm{states}\}$ (step~1). Continuous drives (shown as orange waves) with equal Rabi frequency $\W_1=\W_2=\W$ are then applied on the two qubits for time $t\,\in\,\{t_q\}$, with $q\,\in\,\{1,2,\ldots,N_\mrm{times}\}$, during which S evolves under the influence of the noise spectra evaluated at $\pm\W$, $\{S_{jk}(\pm\W)\}$ with $j,k\,\in\,\{1,2\}$ (step~2). After this evolution time, a projective measurement of a two-qubit observable $O\,\in\,\{O_r\}$, with $r\,\in\,\{1,2,\ldots,N_\mrm{obs}\}$ is performed (step~3). For a given combination $\al$ of initial state $\rho_s$, evolution time $t_q$, and observable $O_r$, this cycle is repeated $M$ times, and a sample mean $\overline O_\al$  of all outcomes $O_\al^{(m)}$ is obtained [Eq. \eq{eqnSampleMeanDef}], where $m$ labels the outcome of the projective measurement for each cycle. The Rabi frequency $\W$ is swept to gather experimental observations for all frequencies at which $S_{jk}(\pm\W)$ will be sampled.
{\bf (b)} Schematic of the reconstruction procedure of the two-qubit spectra from the data produced in (a). An initial value of $\W$ is chosen, for which an initial guess of the spectrum vector $\mvec S(\W)$ is assumed. This guess is fed into a non-linear regression algorithm to find the value of $\mvec S(\W)$ that globally minimizes the discrepancy between the measured sample means $\overline O_\al$ (blue dots with error bars; note, in the first plot, the occurrence of an ``outlier'') and the corresponding expectation values $\langle O_\al \rangle_{\mvec S}$ obtained by numerically solving Eq.~\eq{eqnReduced} for all chosen combinations $\al$ of initial state, evolution time, and observable. An estimate $\hat{\mvec S}(\W^n)$ of $\mvec S(\W^n)$ at the $n$-th Rabi frequency $\W^n$ is obtained following Eq.~\eq{eqnSest} (solid red lines). The latter is then used as the initial guess for the reconstruction at the next frequency $\W^{n+1}$. The procedure is repeated until $\mvec S(\W)$ is reconstructed over all frequencies of interest. 
\label{fig:protocol}}
\end{figure}

Figure~\ref{fig:protocol}(a) illustrates the control and measurement cycle employed to gather experimental data from which $\mvec S(\W)$ is  reconstructed at a given frequency $\W$. 
At time $t=0$, the system is prepared in $\rho(0)=\rho_s$, where $s\in\{1,2,\ldots,N_\mrm{states}\}$ labels elements of an arbitrary set $\{\rho_s\}$ of two-qubit initial states.

Continuous drives resonant with each qubit are then applied with Rabi frequency $\W$, so that the evolution of the two-qubit system is approximately given by the solution to Eq.~\eq{eqnReduced}. Multiple evolution times $t\,\in\,\{t_q\}$, with $q\,\in\,\{1,2,\ldots,N_\mrm{times}\}$, are considered, after which projective measurements of a system's observable $O\,\in\,\{O_r\}$ are performed, with $r\,\in\,\{1,2,\ldots,N_\mrm{obs}\}$. Though we need not specify the initial states and observables at this stage, in the experiment presented here we will consider initial product states in the spin-locking basis, namely, $\rho_s=\proj{\y_s}$, with $\ket{\y_s}\,\in\,$ $\{\ket{{+x},{+x}}$, $\ket{{+x},{-x}}$, $\ket{{-x},{+x}}$, $\ket{{-x},{-x}}\}$, and measurements of products of Pauli operators, for which $O_r=\tau^{\ell_1}_1\otimes\tau^{\ell_2}_2$, with $\ell_1,\ell_2\,\in\,\{0,x,y,z\}$, and where $\tau^0_j\equiv\mathbb{I}_j$ is the identity operator for qubit $j$. These initial states and observables are accessible through simultaneous preparation and measurement of each qubit, and thus using purely local resources.

Figure~\ref{fig:protocol}(b) illustrates the procedure by which experimental observations resulting from the control and measurement cycle described above are used to reconstruct the spectrum vector $\mvec S(\W)$. To simplify the notation, we label all combinations of initial states $\rho_s$, evolution times $t_q$, and observables $O_r$ using a single collective index $\al\,\in\,\{1,2,\dots,d\}$, where $d=N_\mrm{states}\times N_\mrm{times}\times N_\mrm{obs}$. In addition, for each $\al$, we consider sample means $\overline O_\al$ of all outcomes $O_\al^{(m)}$ of projective measurements $m\,\in\;\{1,2,\ldots,M\}$. These sample means are defined by
\begin{align}
\overline{O}_\al\equiv\frac1M\sum_{m=1}^M O_\al^{(m)},	
\label{eqnSampleMeanDef}
\end{align}
where $M$ is the total number of projective measurements, which we take to be the same for each $\al$ for simplicity. It will also be convenient to collect all such sample means measured experimentally for a given Rabi frequency into a single \emph{observation vector}, $\overline{\mvec O}\equiv (\overline O_1, \overline O_2,\ldots, \overline O_d)^T$.

In the asymptotic limit of a large number of projective measurements, $M\rightarrow\infty$, the sample means $\overline O_\al$ converge to their expectation values $\mean{O_\al}_{\mvec S}$ by the weak law of large numbers, $\overline O_\al\rightarrow \mean{O_\al}_{\mvec S}$. These expectation values are determined by $\mvec S$ through
\begin{align}
 \mean{O_\al}_{\mvec S}\equiv\Tr[O_{r_\al}\left.\rho(t_{q_\al})\right|_{\mvec S,s_\al}],    \label{eqnExpectation}
\end{align}
where $\left.\rho(t_{q_\al})\right|_{\mvec S,s_\al}$ is the solution of Eq.~\eq{eqnReduced}, for noise spectra $\mvec S$ and initial state $\rho_{s_\al}$ at time $t_{q_\al}$. Hence, for finite $M$, we estimate the spectra by finding the value of $\mvec S$ that minimizes the deviation between the data $\overline{O}_\al$ and the predictions of the model $\mean{O_\al}_{\mvec S}$. Formally, we define our estimator of $\mvec S$ for a particular Rabi frequency $\W$ as
\begin{align}
\hat{\mvec S}\equiv\argmin_{\mvec S} \sum_{\al=1}^d\ld(z_{\mvec S,\al}).
\label{eqnSest}
\end{align}
Here, $\lambda(z)$ is a \emph{loss function} that penalizes deviations between the model and the data, which are quantified by the normalized \emph{residuals} $z_{\mvec S,\al}\equiv(\overline O_\al-\mean{O_\al}_{\mvec S})/\overline \s_\al$, where $\overline{\s}_\al^2\equiv\mrm{var}(\overline O_\al)$. Throughout this text, we use hats to denote estimators. Estimators like Eq.~\eq{eqnSest}, which minimize a total cost function, are called \emph{M-estimators}~\cite{huber1981robust}.

The most natural choice of loss function is arguably the quadratic function $\lambda(z)=z^2/2$, since Eq.~\eq{eqnSest} then reduces to a simple weighted least-squares estimation. In addition to being well-suited for numerical optimization, the weighted least-squares estimate is statistically well motivated when the probability distribution of $\overline{\mvec O}$ is Gaussian. Indeed, in this case, weighted least-squares optimization can be derived from maximum-likelihood estimation of $\mvec S$, and can be shown to be asymptotically efficient, that is, the variance of $\mvec{\hat S}$ achieves the Cram\'er-Rao bound on precision~\cite{casella2002statistical}.

In practice, however, the statistics of sample means $\overline{\mvec O}$ may not be perfectly described by the expected Gaussian probability distribution, compromising the asymptotic efficiency of $\mvec{\hat S}$ under weighted least-squares estimation. In particular, experimental data is often contaminated by \emph{outliers}: data points that do not follow the probability distribution of the majority, for example because of isolated experimental errors. Because of its quadratic dependence on the residuals, weighted least-squares estimation notoriously gives excessive weight to outliers that are distant from normal observations, namely, for which $z_{\mvec S,\al}\gg1$. This can make weighted least-squares estimators inefficient, causing estimates to wander very far away from their expected behavior (as will be seen experimentally in Sec.~\ref{secValidation} and Appendix~\ref{secHuber}), or, in the worst case, leading to catastrophic divergent behavior of the variance of the estimator. Overcoming these limitations motivates the use of \emph{robust} estimators, whose performance is not significantly impaired by the presence of outliers~\cite{casella2002statistical,huber1981robust}. A prevalent way to achieve robust estimation is to employ the \emph{Huber loss function}, a mixture of weighted least-squares estimation with mean absolute error minimization defined by 
\begin{align}
 \ld(z)\equiv \left\{\begin{array}{ll}
                \frac12 z^2	& \mbox{if }|z|\leq\dt_0,\\
                \dt_0(|z|-\frac12\dt_0)	&\mbox{otherwise,}
               \end{array}	
\label{eqnHuber} \right.
\end{align}
where $\dt_0\,\in\,[0,\infty[$ is a tuning parameter that controls the mixing. For $|z|\leq \dt_0$, the Huber loss function provides the statistical and numerical efficiency of weighted least squares, while residuals with $|z|> \dt_0$ only contribute to the total cost through their absolute value. This  avoids overweighting outliers in the observations and favors robustness of the estimator, as desired.

\section{Correlated noise engineering and validation via circuit QED}
\label{secDemo}

\begin{figure}[t]
\includegraphics[width=8cm]{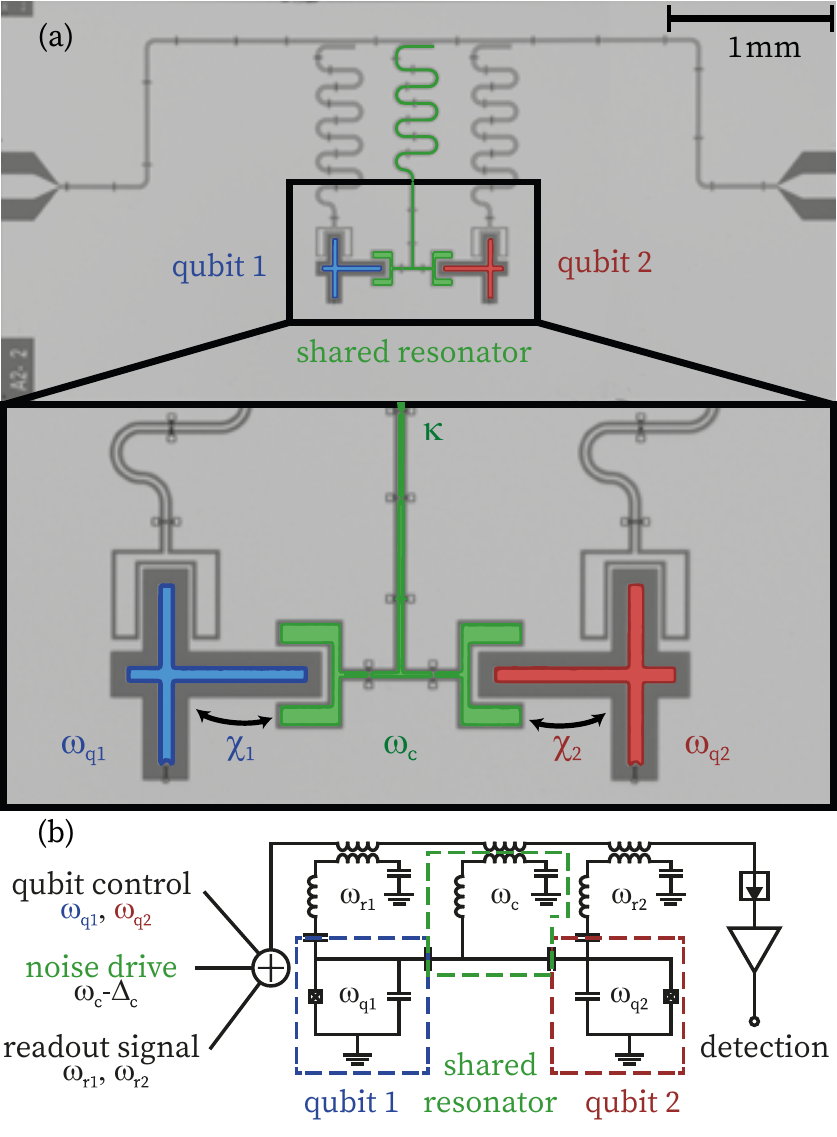}
 \caption{{\em Details of experimental cQED device.} 
 \textbf{(a)} Optical micrograph of the sample. The two qubits (blue, red) are capacitively coupled to a common $\lambda/4$-resonator (green) that we use to create a shared noise source. Each qubit is also coupled to an individual resonator that we use for readout. \textbf{(b)} Simplified circuit diagram of our setup.  
 \label{figMicrograph}}
\end{figure}

In the experiment we will present, we employ the superconducting circuit shown in Fig.~\ref{figMicrograph}(a) as a test-bed for two-qubit spectroscopy based on the spin-locking technique described in Sec.~\ref{secMethodology}. The two qubits are encoded by the lowest energy levels of a pair of transmons dispersively coupled to a common bath formed by a mode of a common resonator, which is brought to a steady-state population under the competing action of a constant applied microwave drive and photon emission into an external environment. When the coupling between the qubits and this bath is sufficiently weak, and the evolution time is sufficiently long, the common bath may be traced out following the derivation presented in Appendix~\ref{secTCLphoton}, and the spin-locking experiment may be described by Eq.~\eq{eqnReduced} for evolution of the two-qubit system under the photon shot-noise spectra~\cite{clerk2010introduction,yan2018distinguishing}:
\begin{align}
S_{jk}(\w)=\chi_j\chi_k\overline{n}\frac{\kp}{(\w+\D_\mrm c)^2+(\kp/2)^2}, 
\label{eqnSpectrumShotNoise}
\end{align}
with $j,k \in \{1,2\}$. Here, $\chi_j$ is the strength of the dispersive coupling for qubit $j$, $\kappa$ quantifies the resonator damping rate, $\D_\mrm c\equiv \w_\mrm c-\w_\mrm d$ is the detuning between the drive frequency $\w_\mrm d$ and the bare frequency of the common resonator $\w_\mrm c$, and $\overline n$ is the average photon number in the steady state of the resonator. While the parameters $\kp$, $\chi_1$, and $\chi_2$ are set by design during device fabrication, both the amplitude and asymmetry of the spectra around zero frequency may be tuned \textit{in situ} via $\overline n$ and $\D_\mrm c$ by varying the strength and detuning of the drive applied onto the common resonator, granting us the capability to produce spatiotemporally correlated noise with an engineered spectrum. 

To validate this capability, in this section we demonstrate the presence of noise correlations consistent with photon shot noise. We first use a Ramsey interferometry technique to witness spatial correlations by measuring a two-qubit correlation function in a free-evolution setting. Fitting the results to the solution of the quantum-optical ME describing the joint evolution of the qubits and the resonator enables us to measure the parameters $\chi_1$, $\chi_2$, and $\kappa$ entering Eq.~\eq{eqnSpectrumShotNoise}. We then use the spin-locking technique to measure noise correlations in a frequency-sensitive fashion and demonstrate experimental control over the amplitude and asymmetry of the engineered noise spectra through both  $\overline n$ and $\D_\mrm c$.

\subsection{Ramsey interferometry}
\label{secRamsey}

\begin{figure*}
 \begin{center}
  \includegraphics[width=18.cm]{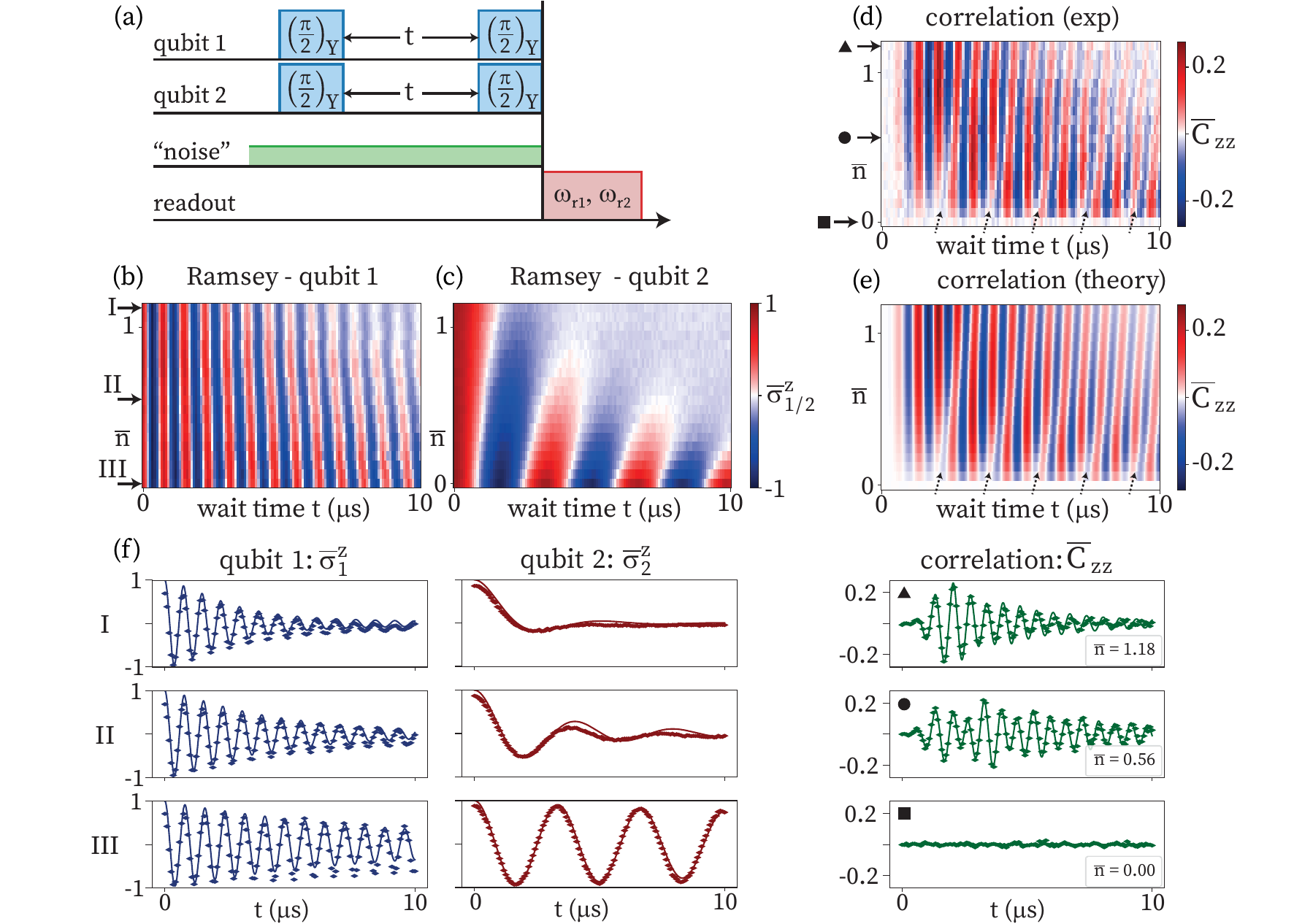}
 \end{center}
 \vspace{-3mm}
 \caption{{\em Ramsey interferometry experiments.}
 \textbf{(a)} Ramsey sequence applied simultaneously to the two qubits, with $\frac\pi2$-pulses around the $y$-axis (blue boxes). A coherent drive (the ``noise'' tone, shown in green) is turned on well before the first pair of $\frac\pi2$-pulses, to bring the common resonator into a steady state with finite population, thus implementing a stationary source of engineered correlated noise. We read out both qubits simultaneously through their individual resonators, using drives at frequencies $\w_\mrm{r1}$ and $\w_\mrm{r2}$ (red box), thus acquiring a series of $\pm 1$ outcomes for each qubit.
\textbf{(b)-(c)} Sample mean of measured single-qubit observables $\overline{\sigma^z_{1,2}}$ for different wait times $t$ and different injected photon numbers $\overline{n}$ in the common cavity. The frequency of the observed Ramsey fringes varies with increasing $\overline{n}$ due to the dispersive shift $2\chi_j\overline{n}$ of the qubit frequencies (tilt of the vertical red and blue lines). For higher $\overline{n}$, we see a rapid dephasing of the qubit states due to the added photon shot noise (blurring of the features in the top right corners). 
\textbf{(d)} Measurement of the correlation $\overline C_{zz}=\overline{\sigma_1^z \sigma_2^z} -\overline{\sigma_1^z}\;\overline{\sigma_2^z}$.
\textbf{(e)} Solution of the ME describing the qubits coupled to the common resonator [Eq. (\ref{eqnMERamsey})], in which parameters $\kp$, $\chi_j$, $\D_{\mrm q j}$, and $\g_{\phi j}$ (see Table~\ref{tabSystemParams}) are obtained by nonlinear regression of the simulation outcomes to the experimental data. 
\textbf{(f)} Comparison of measured ({\tiny $\blacklozenge$}) and simulated (---) values using fitted parameters. Linecuts at  $\overline{n}\in\{0,0.56,1.18\}$ (marked by black arrows in (b) and (d)).
\label{figRamsey}}
\end{figure*}

As mentioned, we create a source of correlated photon shot noise by applying a coherent signal to the common resonator. We then perform simultaneous Ramsey sequences by applying a pair of $\tfrac{\pi}{2}$-pulses on each qubit [Fig.~\ref{figRamsey}(a)]. The qubit drives are detuned from the qubit frequencies by $\D_{\mrm q j}=\omega_{\mrm q j}-\omega_{\mrm d j}$, with $\Delta_{\mrm q 1}/2\pi= -1.26$\,MHz and $\Delta_{\mrm q 2}/2\pi= 0.3$\,MHz. To observe Ramsey fringes, we vary the wait time $t$ between the pulses applied on each qubit, as well as the injected photon number $\overline n$ in the common resonator, which was calibrated in advance by varying the power of the noise drive and observing the frequency shift of the qubit due to the increasing resonator population \cite{macklin2015near}. After each Ramsey sequence, we perform simultaneous single-shot dispersive readout of both qubits in the $\s^z_j$-eigenbasis through their individual resonators, and reinitialize the qubits in the ground state by letting them relax for 500\,$\mu$s, a time much longer than the observed qubit relaxation time $T_1$. During the readout we turn off the otherwise continuous noise drive to improve the single-shot readout fidelity. Taking the sample means $\overline{\s^z_1}$ and $\overline{\s^z_2}$ of the resulting readout outcomes for each qubit [Eq.~\eq{eqnSampleMeanDef}], we obtain the Ramsey fringes shown by the color schemes in Fig.~\ref{figRamsey}(b) and Fig.~\ref{figRamsey}(c), respectively. To access the correlation induced by the shared noise source, we then calculate the sample mean of measurement outcomes of the two-qubit observable $\overline C_{zz}\equiv\overline{\sigma_1^z \sigma_2^z} -\overline{\sigma_1^z}\;\overline{\sigma_2^z}$, where we subtract the product of the single-qubit sample means to remove any potential systematic bias in the readout outcomes.

The non-zero values of $\overline C_{zz}$ shown in Fig.~\ref{figRamsey}(d) clearly indicate the presence of correlated noise in our two-qubit system. To demonstrate that this correlated noise indeed comes from the photons injected in the cavity, we compare experimental results with the numerical solution of the quantum-optical ME~\cite{walls2008quantum}:
\begin{align}
 \dot\rho_\mrm{QC}(t)&=-i\commut{H_\mrm R}{\rho_\mrm{QC}(t)}+\kp\mathcal D[a]\rho_\mrm{QC}(t)	\notag\\
  &\qquad+\mathcal L_x^\mrm R\rho_\mrm{QC}(t)+\mathcal L_z^\mrm R\rho_\mrm{QC}(t). 
  \label{eqnMERamsey}
\end{align}
Here, $\rho_\mrm{QC}(t)$ denotes the joint density matrix for the two qubits plus the resonator mode, which evolves under a dispersive Hamiltonian of the form~\cite{blais2004cavity,gambetta2008quantum}
\begin{align*}
 H_\mrm R &=\sum_j\frac{\D_{\mrm q j}}2\s^z_j+\D_\mrm c a^\dag a+\veps(a+a^\dag)+\sum_j\chi_ja^\dag a\s^z_j,	
 \label{eqnHRamsey}
\end{align*}
with $\varepsilon$ being the amplitude of the drive applied on the resonator. In Eq.~\eq{eqnMERamsey}, the Lindblad superoperators $\mathcal L_x^\mrm R$ and $\mathcal L_z^\mrm R$ are defined by
\begin{align}
 \mathcal L_x^\mrm R\rho\equiv\sum_j\G_{1,j}\mathcal D[\s^-_j]\rho,
 \hspace{5mm}
 \mathcal L_z^\mrm R\rho\equiv\sum_j\frac{\g_{\phi j}}{2}\mathcal D[\s^z_j]\rho,
\end{align}
and describe the effect of noise coupling to qubit $j$ through $\s^x_j$ and $\s^z_j$, leading to relaxation and dephasing at rates $\G_{1,j}=1/T_1^{(j)}$ and $\g_{\phi j}$, respectively. In Eq.~\eq{eqnMERamsey}, $\mathcal L^\mrm R_z$ phenomenologically describes any uncorrelated source of noise that may couple to $\s^z_j$ in addition to photon shot noise.

Solving Eq.~\eq{eqnMERamsey} using standard numerical packages~\cite{johansson2012qutip}, we calculate the time-dependent expectation value $\mean{\s^z_1\s^z_2(t)}-\mean{\s^z_1(t)}\mean{\s^z_2(t)}$ after two instantaneous Ramsey pulses, and estimate relevant physical parameters by fitting to the measurements of $\overline C_{zz}$ displayed in Fig.~\ref{figRamsey}(d). Through this procedure, we obtain the values of $\kp$, $\chi_1$, $\chi_2$, $\D_{\mrm q 1}$, $\D_{\mrm q 2}$, $\g_{\phi 1}$, and $\g_{\phi_2}$ collected in Table~\ref{tabSystemParams}. Throughout the simulations, we take $\D_\mrm c=0$, and use $\G_{1,j}$ measured independently from relaxation experiments in the absence of injected photons.  The results of the simulations using the fitted parameters are displayed in Fig.~\ref{figRamsey}(e). The remarkable quantitative agreement between simulation and experiment obtained here [see also the linecuts in Fig.~\ref{figRamsey}(f) for representative $\overline{n}$ values] provides strong evidence that the measured correlations arise from our engineered source of photon shot noise.

\begin{table}
 \begin{tabular}{l|ccccc}
  \hline\hline
&~~~qubit 1 & &qubit 2 &&common resonator     \\
\hline
$\omega_{}/2\pi$\,(GHz)~& 3.483 &&   4.600 && 7.471\\
$T_1$\,($\mu$s)~& 87 && 54  &&   \\
$T_2^{\mathrm{echo}}$\,($\mu$s)~& 54 && 68  &&  \\
$\kappa_{\mathrm{}}/2\pi$\,(kHz)~&  && &&198   \\
$\chi_{\mathrm{j}}/2\pi$\,(kHz)~& -29.1 && -59.5  &&  \\
$\Delta_{\mrm q j}/2\pi$\,(kHz)~& -1265 &~~~& 299 &~~~& \\
$\gamma_{\phi_j}$\,($\times10^3$rad/s)~& 87.7 &&31.0 &&  \\
  \hline\hline
 \end{tabular}
 \caption{{\em Sample parameters.} The qubit and resonator frequencies are measured spectroscopically, and the reported relaxation and coherence times are averages of repeated $T_1$-decay and echo measurements over 24h, with $T_2<2T_1$ indicating native sources of dephasing noise. The coupling constants $\kappa$ and $\chi_j$, the qubit detunings $\D_{\mrm q j}$ and the native dephasing rates $\g_{\f_j}$, are results of the fits to the Ramsey measurements (Fig.~\ref{figRamsey}). These values are found to agree within error bars with those obtained from independent measurements.
\label{tabSystemParams}}
\end{table}

Several important features of this Ramsey interferometry experiment may be understood qualitatively. First, $\overline C_{zz}$ vanishes when the noise drive is off ($\overline{n}=0$). That is, having (approximately) no photons entering or leaving the common microwave cavity turns off the interaction responsible for a non-zero correlation between the two qubits, as desired. Second, $\overline C_{zz}$ becomes ``blurry'' in the top right corner of Fig.~\ref{figRamsey}(d) for high $\overline{n}$ and $t$. As we increase the amplitude of the photon shot noise, we induce correlated dephasing, implying that more photons enter and leave the common resonator, each one causing the qubits to simultaneously dephase as it takes with it information about the phase of the qubits. At a high enough photon number, the dephasing rate is the dominant source of decoherence, causing the signal to decrease significantly within the wait time of $10\,\mu$s between the Ramsey pulses. Furthermore, we observe oscillations of $\overline C_{zz}$ at frequencies $|\Delta_{\mrm q1}-\Delta_{\mrm q2}|/2\pi=1.56$\,MHz and $|\Delta_{\mrm q1}+\Delta_{\mrm q2}|/2\pi=0.96$\,MHz. These two overlapping signals cause a beating pattern at their difference frequency 0.6\,MHz, which corresponds to the blurry white lines that appear to enter Figs. \ref{figRamsey}(d)-(e) from the bottom left (indicated by dashed arrows).

\subsection{Demonstration of frequency selectivity from measurements of correlated noise in spin-locking}
\label{secSpinlockingExperiment}

\begin{figure*}
 \begin{center}
  \includegraphics[width=0.97\textwidth]{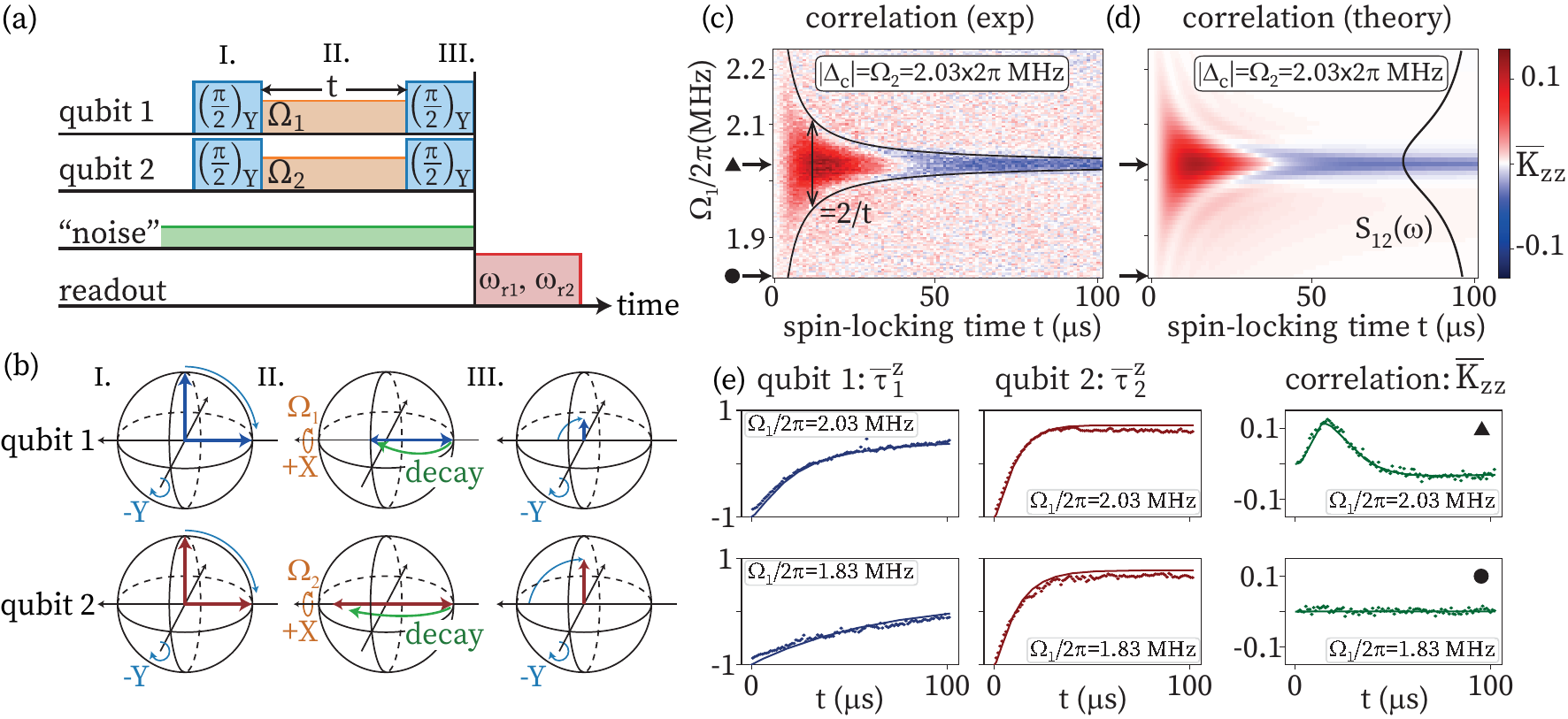}
 \end{center}
 \vspace{-2mm}
 \caption{{\em Spin-locking experiments.}
 \textbf{(a)} Spin-locking sequence applied simultaneously on two qubits with a shared source of engineered noise. 
 \textbf{(b)} Bloch sphere sketch of the sequence in (a) for qubit 1 and qubit 2. The initial $\frac{\pi}{2}$-pulse rotates the qubits from the ground state to $\ket{{-x}}$. The spin-locking drive effectively creates a dressed two-qubit system, with each level splitting being equal to $\Omega_j$. Due to dephasing noise (from both the injected photon shot noise and from weaker native sources),  the dressed qubits decay to their steady-state values along the $x$-axis. The final $\frac{\pi}{2}$-pulse turns the qubits back to the initial ($\s^z_j$) quantization axis to be measured. 
 \textbf{(c)} Measured correlation $\overline{K}_{zz}\equiv\overline{\tau^z_1\tau^z_2}-\overline{\tau^z_1}\;\overline{\tau^z_2}$, where the Pauli matrix $\tau^z_j$ is diagonal in the spin-locking basis \{$\ket{{+x}}_j,\ket{{-x}}_j$\}. The coherent drive creating photon shot noise is detuned by $\Delta_\mrm c/2\pi=-2.03\,$MHz away from the common resonator. The Rabi frequency $\Omega_2$ of the spin-locking drive of qubit 2 is held constant at $\W_2=|\D_\mrm c|$, while $\W_1$ is swept ($y$ axis). The correlation is significant in the region indicated by the solid black line, with width $\sim 2/t$.  
 \textbf{(d)} Numerical solution of the quantum-optical ME [see Eqs.~\eq{eqnHSL} to \eq{eqnLzSL}]. Inset: Qualitative shape of the cross-spectrum $S_{12}$ [Eq.~\eq{eqnSpectrumShotNoise}].  
 \textbf{(e)} Comparison of measured values ({\tiny $\blacklozenge$}) and numerical simulation ({\large{$-$}}) using fitted parameters. The subplots show (from left to right) $\overline{\tau}_1^z$, $\overline{\tau}_2^z$, and $\overline{K}_{zz}$ for $\Omega_1/2\pi\in\{1.83,2.03\}\,$MHz (indicated by black arrows in (c) and (d)). Since $\Omega_2$ is unchanged through the experiment, the decay curves for $\overline{\tau_2^z}$ are roughly the same for both linecuts. The decay curve for $\overline{\tau_1^z}$, however, is steeper for $\Omega_1$ closer to the center of the photon shot noise spectrum (top left plot), demonstrating sensitivity to the noise frequency. The correlation $\overline{K}_{zz}$ roughly equals zero for $\W_1\neq\W_2=|\D_\mrm c|$ (bottom right plot) and shows a clear peak for $\W_1=\W_2=|\D_\mrm c|$ (top right plot). Note the negative steady-state value of the correlation for long spin-locking durations. Here, one of the dressed qubits acts as an effective decay channel for the other, with the coupling between them mediated by the noise.  
\label{figSpin} }
\end{figure*}

In the previous section, we observed the signature of a spatially correlated noise source on two qubits. However, spectral estimation requires the ability to select the frequency at which the noise spectrum is probed. To this end, we use the generalized spin-locking technique we described in Sec.~\ref{secMethodology}.  

Specifically, our protocol is illustrated in Figs.~\ref{figSpin}(a)-(b). Two sets of simultaneous $\tfrac{\pi}{2}$-pulses are applied along the $y$-axis and separated by a time $t$, during which the qubits are driven by constant, resonant drives. The first pair of pulses initializes both qubits in state $\ket{{-x}}$. As explained in Sec.~\ref{secSpinLocking}, the drives define dressed qubits whose eigenstates $\ket{{\pm x}}_j$ are split by the Rabi frequencies $\W_j$. During the time $t$, these dressed qubits undergo a non-unitary evolution due (predominantly) to noise coupling to $\s^z_j$ at angular frequencies $\W_j$. The second pair of $\tfrac{\pi}{2}$-pulses returns the qubits back to the $z$-axis, where they are immediately measured by dispersive readout. We use this sequence to probe correlations in qubit dephasing by sweeping the Rabi frequency $\Omega_1$ across the fixed Rabi frequency $\Omega_2$. The peak frequency of our engineered noise being at $-\D_\mrm c$ [Eq.~\eq{eqnSpectrumShotNoise}], we set $\W_2=|\D_\mrm c|$ to maximize the contribution of shot noise in qubit dynamics. We expect to see a non-zero correlation, $\overline K_{zz}\equiv\overline{\tau^z_1\tau^z_2}-\overline{\tau^z_1}\;\overline{\tau^z_2},$ only if both dressed qubits are sensing the same frequency component of a non-vanishing spectrum ($\W_1\approx\W_2$). This insight is confirmed by the experimental data shown in Fig.~\ref{figSpin}(c), in which $\overline K_{zz}$ is only significant around $\Omega_1=\Omega_2=|\Delta_\mrm c|$, in a frequency region that narrows with the duration $t$ over which the spin-locking drive is applied. This follows from a key feature of the qubit evolution during the spin-locking drive: namely, the terms in the ME that contain the cross-spectra oscillate at $\pm |\W_1-\W_2|$ [see Eq. (\ref{eqnKernelSecular})] and average out for times $\gtrsim 1/|\W_1-\W_2|$. Thus, setting $\W_1=\W_2$ lets us isolate the influence of the cross-spectra in a specific frequency region. 

To verify compatibility of the observed correlation with photon shot noise, we derive the quantum-optical ME for the two qubits and the resonator in the spin-locking frame. This leads to an expression formally identical to Eq.~\eq{eqnMERamsey}, but in which we replace $H_\mrm R\to H_\mrm{SL}$, $\mathcal L_x^\mrm R\to \mathcal L_x^\mrm{SL}$, and $\mathcal L_z^\mrm R\to \mathcal L_z^\mrm{SL}$, where, 
taking $\D_{\mrm q j}=-2\chi_j\overline n$,
\begin{align}
 &H_\mrm{SL}\!=\!\sum_j\!\frac{\W_j}2\tau^z_j+\D_\mrm c a^\dag a+\veps(a+a^\dag)\!-\!\sum_j\!\chi_j(a^\dag a-\overline n)\tau^x_j
 \label{eqnHSL},\\
 &\mathcal L^\mrm{SL}_x\rho=\frac14\sum_j\G_{1,j}\left\{\mathcal D[\tau^z_j]+\mathcal D[\tau^+_j]+\mathcal D[\tau^-_j]\right\}\rho,\\
 &\mathcal L^\mrm{SL}_z\rho=\sum_j\left\{\g^\dwna_j\mathcal D[\tau^-_j] + \g^\upa_j\mathcal D[\tau^+_j]\right\}\rho.	
 \label{eqnLzSL}
\end{align}
Above, $\mathcal L^\mrm{SL}_z$ phenomenologically describes uncorrelated sources of noise coupling to $\s^z_j$ in addition to photon shot noise, in the spin-locking frame. In the ME simulations, we take values of $\chi_j$ and $\Gamma_{1,j}=1/T_1^{(j)}$ given in Table~\ref{tabSystemParams}. In addition, we take the bare cavity drive detuning to be $\D_\mrm c=\D_\mrm c^{(00)}+\chi_1+\chi_2$, where $\D_\mrm c^{(00)}/2\pi=-1.95$~MHz is the cavity drive detuning obtained from experimental measurements of the Stark-shifted resonator transmission peak when both qubits are in their ground state. This yields $\D_\mrm c/2\pi\approx -2.03$~MHz; the remaining parameters are then obtained by fitting the solution of the ME to the experimental data shown in Fig.~\ref{figSpin}(c), leading to $\overline n\approx0.154$, $\gamma^\upa_1\approx2\times10^3$~rad/s, $\gamma^\dwna_1\approx7\times10^3$~rad/s, $\gamma^\upa_2\approx9\times10^3$~rad/s, and $\gamma^\dwna_2\approx14\times10^3$~rad/s. The resulting correlation, shown in Fig.~\ref{figSpin}(d), displays strong quantitative agreement with experimental observations. Though the phenomenological decay rates $\g^\upa_j$ and $\g^\dwna_j$ are not entirely negligible, they lead to decay on a timescale $\gtrsim 100\;\mu$s, while the dynamics due to photon shot noise occur on a shorter timescale, $\lesssim 50$~$\mu$s. This confirms that our engineered noise source predominantly drives the dynamics in this experiment, as intended. 

We finally discuss an additional intriguing feature in the results of the experiment. While the correlation starts out at zero [see Fig.~\ref{figSpin}(e)] when we initiate the spin-locking drive, and then quickly rises to a maximum of about 0.13, $\overline K_{zz}$ ultimately decays over several tens of $\mu$s to a negative value of about $-0.03$. Numerical simulations predict that, despite the presence of intrinsic qubit decay sources (non-zero $\G_{1,j}$, $\g^\upa_j$ and $\g^\dwna_j$), the system subsequently reaches a steady state with $\overline K_{zz}\approx-0.025$. While this negative correlation value may seem puzzling at first, it may be readily understood from the steady state obtained numerically, which contains significant coherences between $\ket{{+x},{-x}}$ and $\ket{{-x},{+x}}$: in turn, this is due to the exchange of dressed qubit excitations mediated by the common source of photon noise.

\section{Validation of two-qubit quantum noise spectroscopy} 
\label{secValidation}

In Sec.~\ref{secDemo}, we used fits to experimental data to estimate the parameters entering the two-qubit spectra of photon shot noise from a coherently-driven common resonator mode, Eq.~\eq{eqnSpectrumShotNoise}. We now use this information to achieve our key objective: validating the spectroscopy protocol presented in Sec.~\ref{secMethodology}, by adapting it to the non-idealities specific to our circuit-QED platform and comparing the resulting experimental spectrum reconstructions with ideal spectra.

\subsection{Experimental non-idealities} 
\label{secNonIdealities}

To produce engineered noise with known spectra, we use a microwave tone to give $\overline n=0.127$. Setting the Stark-shifted resonator-drive detuning with both qubits in their ground state to $\D_\mrm c^{(00)}/2\pi=2.05$~MHz results in a Lorentzian spectrum peaked at angular frequency $-\D_\mrm c$, with $\Delta_\mrm c/2\pi=\D_\mrm c^{(00)}+\chi_1+\chi_2=2\pi\times1.961$~MHz. To produce the experimental data needed to reconstruct these spectra, we apply the two-qubit spin-locking sequence illustrated in Fig.~\ref{fig:protocol}(a) and Fig.~\ref{figSpin}(a)-(b), by letting $\W_1\approx\W_2\approx\W\equiv(\W_1+\W_2)/2$ to have both qubits sample the spectrum at $\w=\W$, thereby maximizing the sensitivity to the noise spatial correlations. We perform a total of $26$ spin-locking experiments, between which the Rabi frequency $\W/2\pi$ is swept through $26$ values uniformly distributed to probe the Lorentzian peak from $-2.2$ to $-1.8$~MHz, along with the corresponding positive frequencies from $1.8$ to $2.2$~MHz. In the spin-locking experiments, we use the $4$ initial states, $11$ observables, and $26$ evolution times given in Table~\ref{tabExample}. To obtain the sample means of the observables, we average over $M=10^4$ simultaneous projective measurements of all $9$ combinations of Pauli matrices $\tau^{\ell_1}_1$ and $\tau^{\ell_2}_2$, $\ell_1,\,\ell_2\,\in\,\{x,y,z\}$, thus performing two-qubit state tomography for each data point~\cite{cramer2012algorithmic}. Separate numerical simulations indicate that the spectra may be accurately reconstructed with fewer initial states and observables for sufficiently high-quality data; nevertheless, we find that full tomography with the $4$ initial states of Table~\ref{tabExample} is more useful in practice, enabling us to diagnose and fix experimental issues such as imperfect calibration of the measurements.
\begin{table}
\begin{tabular}{ll}
  \hline\hline
  Initial states, $\ket{{\y_s}}$	& \quad $\ket{{+x},{+x}}$, $\ket{{+x},{-x}}$,\\   & \quad $\ket{{-x},{+x}}$, $\ket{{-x},{-x}}$.\\
  Observables, $O_r$		& \quad $\tau^z_1$, $\tau^z_2$, $\{K_{\ell_1\ell_2}\},\ell_1,\ell_2\,\in\,\{x,y,z\}$.\\
  Evolution times, $t_q$ 
    & \quad 1, 3, 5, \ldots, 11, 16,\\
 ($\mu$s)	   & \quad 21, 26, \ldots, 71, 81, \ldots, 151.\\
  \hline\hline
\end{tabular}
\caption{{\em Spin-locking control and measurement settings for noise spectroscopy.} 
The two-qubit observables are given by  $K_{\ell_1\ell_2}=\tau^{\ell_1}_1\tau^{\ell_2}_2-\overline\tau^{\ell_1}_1\overline\tau^{\ell_2}_2$, where $\{\tau^\ell_j\},\;\ell\;\in\;\{x,y,z\}$, is the set of Pauli matrices for the dressed qubit $j$ and the bar indicates a sample mean over $M=10^4$ projective measurements.	
\label{tabExample}}
\end{table}

After completing all the spin-locking experiments, we condense the tomography data into sample means of projective measurements of single-qubit observables $\tau^z_j,\;j\;\in\;\{1,2\}$, and two-qubit observables of the form $K_{\ell_1\ell_2}\equiv \tau^{\ell_1}_1\tau^{\ell_2}_2-\overline{\tau}^{\ell_1}_1\overline{\tau}^{\ell_2}_2,$ $\ell_1,\ell_2\,\in\,\{x,y,z\}$, along with corresponding standard deviations. To accurately reconstruct the engineered noise spectra, we find that the ME derived in Sec.~\ref{secSpinLocking} under pure-dephasing noise (in the lab frame) and used for non-linear regression in the procedure described in Sec.~\ref{secEstimation} must be adapted to two types of non-idealities in our cQED setting:

(i) \emph{Finite Rabi-frequency difference}.-- In Sec.~\ref{secSpinLocking}, we assumed $\W_1=\W_2$ to arrive at the reduced ME, Eq.~\eq{eqnReduced}. Experimentally, however, we observe that the amplitude of the spin-locking drives can drift over a timescale of a few hours, most plausibly due to drifts in electronics, thus making $\dt\W\equiv\W_1-\W_2\neq0$. As a consequence, in the interaction picture with respect to $H'_\mrm S$ [explicitly given below Eq.~\eq{eqnHR}], terms involving cross-spectra in Eq.~\eq{eqnReduced} oscillate at frequency $\dt\W$, which significantly suppresses their influence over times $\gtrsim 1/\dt\W$ (see Appendix~\ref{secTCLrabiT1}). This biases any estimate of the spectra based on Eq.~\eq{eqnReduced}.

(ii) \emph{Relaxation noise}.-- Although the engineered dephasing mechanism used in this experiment is made dominant by applying a sufficiently strong resonator drive so that $\overline{n}$ is large, superconducting qubits always suffer from significant intrinsic noise coupling to $\s^x_j$. This  leads to $T_1$ relaxation (e.g., from Purcell decay~\cite{houck2008controlling}) in the lab frame. Such noise has distinct dynamical effects that are not captured by Eq.~\eq{eqnReduced}, and thus biases any estimates of $S_{jk}(\w)$ based on Eq.~\eq{eqnReduced} with respect to their true physical value.

In order to simultaneously account for both types of non-idealities, we derive a modified ME for the reduced two-qubit dynamics. Under the approximations described in Appendix~\ref{secTCLrabiT1}, we arrive at
\begin{align}
 &\dot \rho(t)= -\frac i2\left[(\W+\dt\W/2)\tau^z_1+(\W-\dt\W/2)\tau^z_2,\rho(t)\right]\notag\\
 &+\sum_{jk}\mathcal L_{jk}\rho(t)+\frac14\sum_j\G_{1,j}\left(\mathcal D[\tau^z_j]+\mathcal D[\tau^+_j]+\mathcal D[\tau^-_j]\right)\rho(t),	\label{eqnModifiedReduced}
\end{align}
where $\W$ is now defined as the average Rabi frequency, $\W\equiv(\W_1+\W_2)/2$. In addition, in Eq.~\eq{eqnModifiedReduced}, $\{\mathcal L_{jk}\}$ is the usual set of Lindblad superoperators for correlated dephasing noise given by Eq.~\eq{eqnCorrelatedDecay}, and $\G_{1,j}\equiv 1/T_1^{(j)}$, with $T_1^{(j)}$ the longitudinal relaxation time of qubit $j$ in the lab frame (without spin-locking drives). We then apply the spectral reconstruction procedure described in Sec.~\ref{secEstimation}, with the exception that Eq.~\eq{eqnModifiedReduced} is used in lieu of Eq.~\eq{eqnReduced} in the calculation of expectation values involved in the non-linear regression procedure, Eq.~\eq{eqnExpectation}.

In comparison with Eq.~\eq{eqnReduced}, the modified ME in Eq.~\eq{eqnModifiedReduced} involves three additional parameters: $T_1^{(1)}$, $T_1^{(2)}$, and $\dt\W$. While the longitudinal relaxation times are known from prior characterization of the qubits in the lab frame  (Table~\ref{tabSystemParams}), the Rabi frequency differences are due to slow random drifts, and cannot be known in advance. However, $\dt\W$ can be accounted for as an additional unknown parameter in the estimation scheme; formally, we simply replace $\mvec S\to\mvec \q\equiv [\mvec S,\dt\W]$ in Eq.~\eq{eqnSest} and simultaneously estimate $\mvec S$ and $\dt\W$. In this approach, we thus assume that $\dt\Omega$ is a constant parameter for a given reconstruction at target Rabi frequency $\W$, but allow $\dt\W$ to vary between spin-locking experiments aiming to reconstruct spectra about distinct target Rabi frequencies, thus modelling slow, quasi-static Rabi-frequency drifts.

\subsection{Spectral estimation results} 
\label{secEstimationResults}

\begin{figure*}[tbp]
\begin{center}
\includegraphics[width=0.85\textwidth]{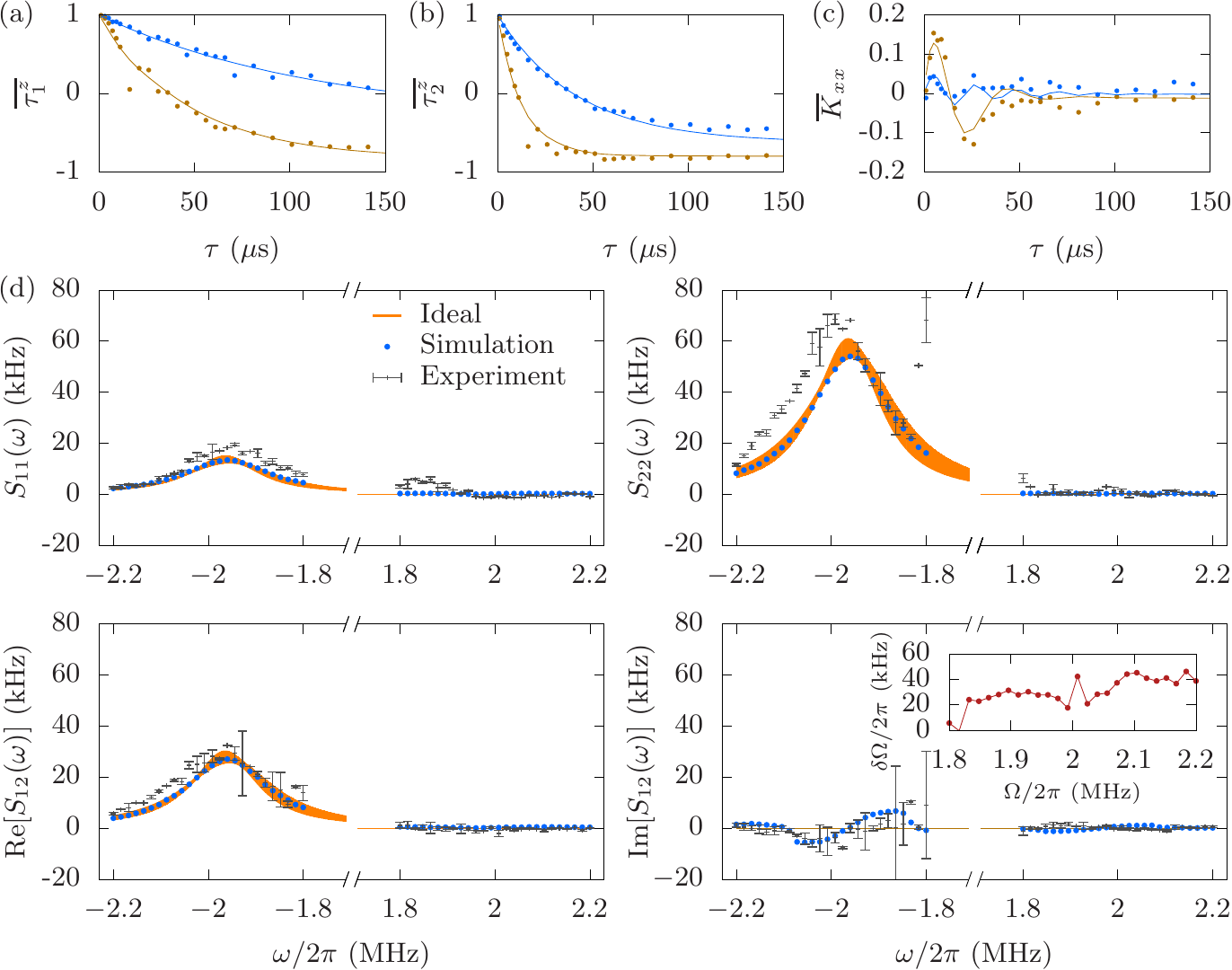}
\end{center}
\vspace{-5mm}
\caption{{\em Two-qubit quantum noise spectroscopy.}
\textbf{(a)-(c)} Representative decay curves from experimental data (dots) and corresponding solution of the reduced ME (solid lines), Eq.~\eq{eqnModifiedReduced}, with estimates for the spectra $S_{jk}(\pm\W)$, $j,k\;\in\;\{1,2\}$, and Rabi-frequency difference $\dt\W=\W_1-\W_2$ obtained by non-linear regression. The sample means are obtained with the two dressed qubits initially prepared in state $\ket{{+x},{+x}}$, and driven at average Rabi frequency $\W=(\W_1+\W_2)/2=2\pi\times 2.184$~MHz (blue dots) or $1.976$~MHz (yellow dots). Using the Huber loss function [Eq.~\eq{eqnHuber}] makes the fitting procedure robust to outliers visible in the data [e.g., the seventh yellow data point in (a)]. 
\textbf{(d)}
Reconstructed two-qubit spectra. Error bars: 95\% confidence intervals for the experimental reconstructions. Shaded orange areas: 95\% confidence intervals associated with the ideal spectra for engineered photon shot noise obtained from Eq.~\eq{eqnSpectrumShotNoise}, with parameters taken from fits described in Sec.~\ref{secDemo}: $\kappa/2\pi=198$~kHz, $\chi_1/2\pi=-29.1$~kHz, $\chi_2/2\pi=-59.5$~kHz, $\D_\mrm c/2\pi =1.961$~MHz, $\overline n=0.127$. Blue dots: numerical simulation of the spectroscopy protocol using the quantum-optical ME in Eqs.~\eq{eqnHSL}--\eq{eqnLzSL}. Inset of the bottom-right panel: Rabi-frequency difference $\dt\W$. 
\label{figSpectrum}}
\end{figure*}

Figure~\ref{figSpectrum} summarizes the results of the spectral reconstruction procedure described above, using a spectrum vector that is uniform across its components, $S_\ell=1$~kHz $\forall \ell$, and $\dt\W=0$, respectively, as initial guesses for the non-linear regression technique we employ. In Fig.~\ref{figSpectrum}(a)-(c), we show examples of experimentally measured decay curves with the dressed qubits initialized in state $\ket{{+x},{+x}}$ for $\W/2\pi=2.184~$MHz (blue dots) and $\W/2\pi =1.976$~MHz (yellow dots) for three of the eleven observables given in Table~\ref{tabExample}. The fitted decay curves are shown by solid lines of corresponding color, and display reasonable agreement with data.  The spectra that follow from this robust estimation procedure are shown in Fig.~\ref{figSpectrum}(d), in which the gray error bars indicate the $95\%$ confidence intervals derived from the asymptotic statistics of M-estimators (Appendix~\ref{secConfidenceIntervals}). 
These reconstructions are compared with shaded orange areas representing $95\%$ confidence intervals for the theoretical shot-noise spectra [Eq.~\eq{eqnSpectrumShotNoise}], using estimates for $\chi_1$, $\chi_2$, $\overline n$, and $\kappa$ obtained as explained in Sec.~\ref{secDemo}. The four experimental reconstructions clearly capture the asymmetry of the two-qubit spectra arising from the non-commuting nature of the noise operator, and qualitatively reproduce the Lorentzian shape associated with photon shot noise. Remarkably, this includes a reconstruction of the cross-spectrum $S_{12}(\w)$ that characterizes spatial correlations of the noise. 

To produce the reconstructions shown in Fig.~\ref{figSpectrum}, we considered all initial states, evolution times and observables given in Table~\ref{tabExample} in a global fit defined by the M-estimator given by Eq.~\eq{eqnSest} with $\mvec S$ replaced by $\mvec \q$. To minimize the total cost function in Eq.~\eq{eqnSest} numerically, we employed the least-squares optimization routine of the SciPy package, which implements a trust-region reflective algorithm that allows the use of arbitrary loss functions~\cite{branch1999subspace}. While the quadratic loss function is the most standard in nonlinear regression, we find that due to the significant presence of outliers in the experimental data, quadratic loss leads to fitted decay curves that can deviate from the large majority of experimental observations. To mitigate this adverse behavior, we thus perform robust estimation using the Huber loss function, Eq.~\eq{eqnHuber}, in which we set the tuning parameter $\dt_0=1$, leading to less noisy reconstructions and significantly better agreement between fitted decay curves and experimental data (see Appendix~\ref{secHuber} for further discussion). This result showcases the advantage of robust estimation strategies over traditional weighted-least squares estimates in a quantum noise spectroscopy context. Indeed, for each of the $26$ Rabi frequencies considered here, the current approach involves, as mentioned, $4$ initial states, $11$ observables, and $26$ evolution times, for a total of $29,744$ data points. For such a large dataset, systematically identifying, explaining, and eliminating outliers in an experimental setting would represent, at best, an extremely tedious and impractical task.

Though the reconstructed and predicted spectra agree within error bars for several frequency values, statistically significant deviations are also observed. In particular, the reconstructed spectra are larger than predicted by an amount $\lesssim 10\;\mrm{kHz}$ for wide ranges of frequencies. To investigate the physical origin of this discrepancy, we first simulate the spectroscopy procedure by exactly solving the coupled evolution of the two qubits and the driven resonator mode using the full quantum-optical ME defined by Eqs.~\eq{eqnHSL}--\eq{eqnLzSL}, setting $\gamma^\upa_j=\gamma^\dwna_j=0\,\forall\, j$ and $\W_1=\W_2=\W$. We then employ the resulting density matrix to calculate the probabilities of all relevant measurement outcomes, and use these probabilities to produce $10,000$ simulated projective measurements for each data point measured experimentally. The resulting decay curves are then fitted using the approach presented in Sec.~\ref{secMethodology}, producing the spectrum estimates illustrated by the blue dots in Fig.~\ref{figSpectrum}. A small discrepancy with the ideal spectra for the same shot-noise parameters then arises because the assumptions used to arrive at the reduced ME for two-qubit evolution are only valid in an approximate sense. In particular, noise may not be sufficiently weak, and the filters for finite evolution time may not be sufficiently narrow for Eq.~\eq{eqnModifiedReduced} to hold exactly. Nevertheless, this discrepancy remains too small to explain the excess noise measured experimentally.

To attempt to explain the observed discrepancies, we invoke non-idealities that, taken together, may help to better understand our experimental results. First, native dephasing noise sources (in addition to engineered shot noise) may couple to the qubits via $\sigma^z_j$. We expect the spectrum of this excess noise to add up with engineered noise in the spectra $S_{jk}(\w)$ appearing in Eq.~\eq{eqnCorrelatedDecay}; such noise would thus also be measured by our protocol. The presence of intrinsic noise would be consistent with the numerical fits performed in Sec.~\ref{secSpinlockingExperiment}, which yielded decay rates between $2\times10^3$~rad/s and $14\times10^3$~rad/s, phenomenologically accounting for spatially uncorrelated noise along $\s^z_j$. In addition, characterization of the qubits without engineered noise led to $T_2$ values significantly shorter than $2T_1$ (see Table~\ref{tabSystemParams}). Native dephasing noise may arise from a combination of residual thermal photons in the readout resonators~\cite{yan2018distinguishing} with sources of noise intrinsic to transmon qubits, such as two-level fluctuators. In particular, the $E_\mrm J/E_\mrm C$ ratios in our sample, which determine the sensitivity of the transmon frequency to charge noise, were $28$ for qubit 1 and $45$ for qubit 2. These are well below the value of $50$ normally associated with transmon devices  \cite{koch2007charge,schreier2008suppressing}.

We also observe that Rabi frequencies significantly drift during the $24$ hours needed to complete all the spin-locking experiments. Indeed, the inset of the bottom-right panel of Fig.~\ref{figSpectrum} shows that a Rabi frequency difference $\dt\Omega/2\pi\sim50\;\mrm{kHz}$ builds up as the target value of $\W/2\pi$ is swept from $1.8$ MHz to $2.2$ MHz, most plausibly due to drifts in electronics. It is thus very likely that the average Rabi frequency $\W/2\pi\equiv(\W_1+\W_2)/2$ defining the frequency at which the spectrum is reconstructed drifts by a similar quantity. In addition, both $T_1$ and $T_2$ are known to fluctuate significantly under noise processes that naturally occur in superconducting qubits~\cite{gustavsson2016suppressing,yan2016flux}. These combined effects may thus explain the apparent shift of some reconstructed spectrum values away from the predicted Lorentzians. 

Finally, the two reconstructed values of $S_{22}(\w)$ at frequencies nearest to $\w/2\pi=-1.8\;\mrm{MHz}$ deviate significantly from predictions. For these Rabi frequencies, we find that the nonlinear regression fails to produce a good fit to the experimental decay curves. We attribute this effect to the weakness of the engineered noise source at these frequencies, leading to a poor signal-to-noise ratio in the measured dynamics that prevents convergence of the nonlinear regression to the correct spectrum values. By artificially down-weighting the residuals associated with two-qubit observables in the loss function, we were able to achieve a more accurate reconstruction of the self-spectra for these frequencies; however, this came at the expense of an inaccurate reconstruction of the cross-spectrum. To achieve simultaneously accurate reconstructions of all spectra far from the peak frequency, it is possible in principle to perform more projective measurements to increase signal-to-noise and thus ease the convergence of the non-linear regression procedure into a physically meaningful global minimum.

\section{Discussion and outlook}

We have proposed a protocol for two-qubit noise spectroscopy and validated it with an engineered source of photon shot noise in a superconducting-qubit architecture. Despite the complexity of the two-qubit dynamics and the resulting estimation problem in the presence of both non-classical self- and cross-correlation spectra, we were able to successfully extend the spin-locking technique previously used on a single qubit to the two-qubit setting. This enabled us to demonstrate what constitutes, to the best of our knowledge, the first experimental reconstruction of a two-qubit noise cross-spectrum.

Our approach offers several advantages over available proposals for two-qubit spectroscopy. Indeed, our continuous-wave protocol avoids the experimental issues arising from long tailored sequences of nearly instantaneous pulses that are typically required in comb-based dynamical-decoupling spectroscopy \cite{paz2017multiqubit}; it does not require two-qubit gates or entangled initial qubit states; and it allows one to probe noise in a MHz frequency scale that is difficult to access with pulsed techniques. In addition, the numerical approach taken here, in which the numerical solution of a ME containing the spectra of interest is fitted to experimental data, offers substantial flexibility in the choice of theoretical model. This enabled us to quantitatively account for non-idealities of our experimental platform in a straightforward manner, by simply modifying the ME to include $T_1$ effects and Rabi-frequency drifts. We expect the simplicity and flexibility of our approach to be instrumental in its application to study correlated noise sources in other state-of-the-art quantum devices -- including nuclear-spin, phonon or charge noise affecting spin qubits in semiconductors or NV centers~\cite{boter2019spatial,kwiatkowski2018decoherence,bennett2013phonon}.

Our work also paves the way to further advances of quantum noise spectroscopy protocols and to the characterization of more general sources of noise. Indeed, applying the same non-linear regression techniques to a different master equation should enable spectral estimation to be extended to multi-axis noise, in a simpler continuous-wave approach than in existing pulsed schemes~\cite{paz-silva2019extending} -- with further extensions possible to a multi-qubit scenario. In addition, as photon shot noise is genuinely non-Gaussian, increasing the strength of the engineered noise studied here should enable one to use non-Gaussian noise spectroscopy methods~\cite{norris2016qubit,sung2019non} to investigate a quantum non-Gaussian environment with non-commuting degrees of freedom.  As noise spectroscopy methods will continue to grow in complexity and generality, more elaborate measurements and larger experimental data sets will also be more likely to contain outliers. We thus envision that robust estimation strategies based on M-estimators, as we demonstrated here, will become increasingly needed and prove useful for quantum sensing applications.

\section*{Acknowledgements}

It is a pleasure to thank to F\'elix Camirand-Lemyre, Gerardo Paz-Silva, and Ioan Pop for fruitful discussions. This research was funded in part by the \textit{U.S. Army Research Office} Grant No. W911NF-14-1-0682 (to L.V., S.G. and W.D.O.); by the \textit{Department of Defense} via MIT Lincoln Laboratory under Air Force Contract No. FA8721-05-C-0002 (to W.D.O.); and by the \textit{National Science Foundation} Grant No. PHY-1720311(to S.G.). F.B. also acknowledges financial support by the \textit{Fonds de Recherche du Qu\'ebec -- Nature et Technologies}.

\appendix

\section{Derivation of the reduced master equations 
\label{secTCL}}

In this Appendix, we derive the reduced MEs presented in the main text, Eq.~\eq{eqnReduced} and \eq{eqnModifiedReduced}, using a standard time-convolutionles
s (TCL) approach~\cite{breuer2002theory}. We first derive the ME in the ideal setting in which noise only couples to $\s^z_j$, $j\in\{1,2\}$. We then evaluate the relevant spectra for the cQED system considered in the main text. Finally, we derive a modified ME which  accounts for non-idealities that are particularly relevant to our experimental setting.

\subsection{Master equation for noise along $\s^z_j$ only 
\label{secTCLideal}}

The TCL formalism enables the systematic derivation of a ME for a system of interest S coupled to a bath B by tracing out the bath degrees of freedom. Though the method presented here is more general, to apply the TCL formalism to the cQED sytem studied experimentally in Sec.~\ref{secDemo} and Sec.~\ref{secValidation}, we will assume that the bath B may be separated into two components: a \emph{central bath} labeled by C, corresponding to a mode of the microwave resonator shared by the two qubits in the main text, and a larger \emph{external bath} labeled by E, corresponding to the environment of the resonator mode in the main text. While the central bath couples to the system, the external bath only couples to the central bath; its effect is to allow non-unitary evolution of the central bath even in the absence of coupling to the system. In our cQED context, this allows us to describe damping of the microwave resonator. Beyond the current setting, such ``structured environments'' can naturally arise, for instance, for qubits coupled to two-level charge fluctuators undergoing incoherent transitions due to an electronic or phononic reservoir~\cite{beaudoin2015microscopic,paladino2014noise}.

More formally, let the system-bath Hamiltonian be 
\begin{align}
 H=H_\mrm S+H_\mrm{SB}+\underbrace{H_\mrm C+H_\mrm{CE}+H_\mrm E}_{\equiv H_B},  \label{eqnHSCE}
\end{align}
where $H_\mrm{SB}\equiv H_\mrm{SC}\equiv \sum_j B_j Q_j$ describes coupling between qubit $j$ and C through the qubit operator $Q_j$ and the central-bath operator $B_j$. Moving to the interaction picture with respect to $H_\mrm S+H_\mrm B$, we have
\begin{align}
 \tilde H_\mrm{SB}(t)=\sum_j \tilde B_j(t) \tilde Q_j(t),	\label{eqnHtilde}
\end{align}
where the transformed operators $\tilde B_j(t)\equiv\eul{i H_\mrm B t}B_j\eul{-i H_\mrm B t}$ and $\tilde Q_j(t)\equiv\eul{i H_\mrm S t}Q_j\eul{-i H_\mrm S t}$. To derive a TCL ME describing the reduced evolution of the qubit system only, we introduce the projection superoperator $\mathcal P$ that projects any density matrix $\rho$ onto the relevant (system) part of the Hilbert space: that is, $\mathcal P\rho\equiv\mrm{Tr}_\mrm B[\rho]\otimes\rho_\mrm B$, where the trace is taken over both central and external baths and $\rho_\mrm B$ is the joint initial state of C and E. A complementary projection superoperator $\mathcal Q$ on the irrelevant part of the density matrix is then also defined by $\mathcal Q\equiv \mathcal I-\mathcal P$, where $\mathcal I$ is the identity superoperator, $\mathcal I \rho\equiv\rho$. As customary, we assume that the density matrix $\rho_\mrm{tot}(0)$ describing the joint initial state of the system and the bath is of the form $\rho_\mrm{tot}(0)=\rho(0)\otimes\rho_B$, where $\rho(0)$ is the initial density matrix of S. Further assuming that coupling between the system and the central bath is sufficiently weak, the TCL ME may be truncated at second order, leading to
\begin{align}
 \dels{t}\mathcal P\tilde\rho_\mrm{tot}(t)=\mathcal K(t)\mathcal P\tilde\rho_\mrm{tot}(t). 
 \label{eqnTCLgeneral}
\end{align}
Here, $\tilde \rho_\mrm{tot}(t)$ denotes the joint interaction-picture density matrix of the system, central and external baths at time $t$, and $\mathcal K(t)$ is the second-order TCL generator given by
\begin{align}
 \mathcal K(t)=\int_{0}^t ds\,\mathcal P\mathcal L(t)\mathcal Q\mathcal L(s)\mathcal P,	
 \label{eqnKorder2}
\end{align}
with $\mathcal L(t)\rho\equiv-i[\tilde H_\mrm{SB}(t),\rho]$ defining the Liouvillian superoperator associated with $\tilde H_\mrm{SB}(t)$. Importantly, Eqs.~\eq{eqnTCLgeneral} and \eq{eqnKorder2} are valid when $\mathcal P\mathcal L(t)\mathcal P=0$ $\forall t$, a property that is satisfied when $\Tr_\mrm B[\tilde B_j(t)\rho_\mrm B]=0$, i.e., for noise with vanishing mean in state $\rho_B$.

Under the above assumption, substituting Eq.~\eq{eqnHtilde} into Eq.~\eq{eqnKorder2} and tracing over both central and external baths produces the integro-differential equation 
\begin{align}
 \hspace{-2.5mm}&\dot{\tilde\rho}(t)\!=\!\sum_{jk}\!\int_0^t\!\!ds\!
  \left[C_{jk}(t,s)\!\!\left(\!\tilde Q_k(s)\tilde\rho(t)\tilde Q_j(t)\!-\!\tilde Q_j(t)\tilde Q_k(s)\tilde\rho(t)\!\right)\right.\notag\\
  &+\left.C_{kj}(s,t)\!\left(\!\tilde Q_j(t)\tilde\rho(t)\tilde Q_k(s)\!-\!\tilde\rho(t)\tilde Q_k(s)\tilde Q_j(t)\!\right)\right],\!\!	\label{eqnIntegroDifferential}
\end{align}
which describes evolution of the reduced density matrix of S, $\tilde \rho(t)\equiv\Tr_\mrm B\tilde\rho_\mathrm{tot}(t)$, in terms of the two-point correlation functions 
\begin{align}
 C_{jk}(t,s)&\equiv\mean{\tilde B_j(t)\tilde B_k(s)}_\mrm B\equiv\Tr_\mrm B[\tilde B_j(t)\tilde B_k(s)\rho_\mrm B]	\label{eqnCorrelation}\\
  &=\Tr_\mrm B[\tilde B_j(t-s)\tilde B_k(0)\rho_\mrm B]\equiv C_{jk}(t-s).	
  \label{eqnStationarity}
\end{align}
Equality between Eq.~\eq{eqnCorrelation} and Eq.~\eq{eqnStationarity} is only respected when noise arising from the central and external baths is stationary, so that correlation functions are invariant under time translations. Noting that $\tilde B_j(t)=\eul{i H_\mrm B t}B_j\eul{-iH_\mrm B t}$ is formally equivalent to the Heisenberg-picture evolution of $B_j$ under the Hamiltonian $H_\mrm B$ of the bath only, $C_{jk}(t,s)$ is independent of the evolution of the system. Stationarity then arises in the following two situations, the second one being the most relevant to this paper: (i) The initial density matrix of the bath commutes with $H_\mrm B$, $[H_\mrm B,\rho_\mrm B]=0$. (ii) Evolution of the reduced density matrix of the central bath in the absence of the system, namely, 
$$\rho_\mrm C(t)\equiv \Tr_\mrm E [\rho_\mrm B(t)]\equiv\Tr_\mrm E[\eul{-i H_\mrm B t}\rho_\mrm B\eul{i H_\mrm B t}],$$ 
is accurately described by a ME of the form $\dot \rho_\mrm C(t)=\mathcal L_\mrm C\rho_\mrm C(t)$, where $\mathcal L_\mrm C$ is a Markovian (Lindblad) superoperator acting on C only. This is the case, for example, when the central and external baths are initially in a product state, $\rho_\mrm B=\rho_\mrm C(0)\otimes\rho_\mrm E(0)$, and C is sufficiently weakly coupled to an external bath  containing enough degrees of freedom for the Born-Markov approximation to hold. The desired correlation function $C_{jk}(t,s)$ is then given by the following expression of multitime averages for evolution under a Markovian ME~\cite{gardiner2000quantum}
\begin{align}
 C_{jk}(t,s)=\left\{\begin{array}{ll}
                     \Tr_\mrm C\left[B_j\eul{\mathcal L_\mrm C(t-s)}B_k\eul{\mathcal L_\mrm C s}\rho_\mrm C(0)\right],	& t\geq s,\\
                     \Tr_\mrm C\left[B_k\eul{\mathcal L_\mrm C(s-t)}\left(\eul{\mathcal L_\mrm C t}\rho_\mrm C(0)\right)B_j\right],	& t<s.
                    \end{array}
\right.	\label{eqnMultitime}
\end{align}
Further assuming that the initial state of the central bath is a steady state of the ME, $\mathcal L_\mrm C\rho_\mrm C(0)=0$, then directly leads to a correlation function respecting Eq.~\eq{eqnStationarity}, and thus to stationary noise.

For stationary noise, Eq.~\eq{eqnIntegroDifferential} can be rewritten in the frequency domain as
\begin{align}
 \dot{\tilde\rho}(t)=\frac1{2\pi}\sum_{jk}\int_{-\infty}^\infty d\w\,\tilde\Upsilon_{jk}(\w,t)\tilde\rho(t),	\label{eqnMEkernel}
\end{align}
where the integration kernel $\tilde\Upsilon_{jk}(\w,t)$ is given by
\begin{widetext}
 \begin{align}
 \tilde\Upsilon_{jk}(\w,t)\rho\equiv\!\!\int_0^t\!d\tau\eul{i\w \tau}\!\!\left\{\!S_{jk}(\w)\!\left[\tilde Q_k(t-\tau)\rho\tilde Q_j(t)-\tilde Q_j(t)\tilde Q_k(t-\tau)\rho\right]
  + S_{kj}(-\w)\!\left[\tilde Q_j(t)\rho\tilde Q_k(t-\tau)-\rho\tilde Q_k(t-\tau)\tilde Q_j(t)\right]\!\right\}\!. 
  \label{eqnKernel}
 \end{align}
 To describe spin-locking, we replace $H_\mrm S\to H'_\mrm S$ and $Q_j\to-\tau^x_j$ in $\tilde Q_j(t)=\eul{i H_\mrm S t}Q_j\eul{-i H_\mrm S t}$, with $H'_\mrm S$ given by Eq.~\eq{eqnHSBspinLocking}, leading to $\tilde Q_j(t)=-(\eul{i\W_j t}\tau^+_j+\mrm{H.c.})$. Substituting this into Eq.~\eq{eqnKernel} above, we obtain terms oscillating at angular frequencies $|\Omega_j-\Omega_k|$ and $\Omega_j+\Omega_k$. Assuming that $(\W_j+\W_k)t_\mrm D\gg 1$, $\forall j,k$, where $t_\mrm D$ is the typical timescale over which qubit observables decay, we neglect terms oscillating with $\Omega_j+\Omega_k$ by invoking a secular approximation~\cite{breuer2002theory}. This leads to the simplified expression 
 \begin{align}
  &\tilde\Upsilon_{jk}(\w,t)\rho \approx  \label{eqnKernelSecular} \\
  &\qquad\: \eul{i\D_{jk}t}\!\left\{\left[S_{jk}(\w)F(\w\!+\!\W_k)\!+\!S_{jk}(-\w)F(\w\!-\!\W_j)\right]\tau^-_k\rho\tau^+_j
    \!\!-\!S_{jk}(\w)F(\w\!+\!\W_k)\tau^+_j\tau^-_k\rho - S_{jk}(-\w)F(\w\!-\!\W_j)\rho\tau^+_j\tau^-_k\right\}	\notag\\
    &\quad+ \eul{-i\D_{jk}t}\!\left\{\left[S_{jk}(\w)F(\w\!-\!\W_k)+S_{jk}(-\w)F(\w\!+\!\W_j)\right]\tau^+_k\rho\tau^-_j
    -S_{jk}(\w)F(\w\!-\!\W_k)\tau^-_j\tau^+_k\rho - S_{jk}(-\w)F(\w\!+\!\W_j)\rho\tau^-_j\tau^+_k\right\},	\notag
 \end{align}
 \end{widetext}
 where $\D_{jk}\equiv\W_j-\W_k$ and $F(\w)\equiv\int_0^tds\,\eul{i\w s}$ is the first-order fundamental filter function for free-induction decay~\cite{paz-silva2014general}. Crucially, for finite time, $F(\w\pm\W_j)$ is peaked around $\w=\mp\W_j$ with a width $\sim 1/t$ and thus acts as a bandpass filter for the noise spectra in the integral over frequencies, Eq.~\eq{eqnMEkernel}. Assuming that all spectra vary negligibly over this passband, we replace $F(\w)$ by its infinite-time limit in the sense of distributions, $\lim_{t\to\infty}F(\w)=\pi\dt(\w)$, which gives
 \begin{align}
  &\dot{\tilde\rho}(t)\approx \sum_{jk}\left(\eul{i\D_{jk}t} \mathcal L^-_{jk} + \eul{-i\D_{jk}t}\mathcal L^+_{jk}\right)\tilde\rho(t),\\
  &\mathcal L^\pm_{jk}\rho\equiv \frac12\left\{\left[S_{jk}(\pm\W_k)+S_{jk}(\pm\W_j)\right]\tau^\pm_k\rho\tau^\mp_j\right.\notag\\
  &\qquad \left.-S_{jk}(\pm\W_k)\tau^\mp_j\tau^\pm_k\rho - S_{jk}(\pm\W_j)\rho\tau^\mp_j\tau^\pm_k\right\}. 	\label{eqnLjkplusminus}
 \end{align}
Moving back to the spin-locking reference frame, which rotates at the qubit drive frequencies [see the discussion above Eq.~\eq{eqnHR}], and taking $\W_1=\W_2$ then results in Eq.~\eq{eqnReduced} in the main text, which provides the theoretical basis of the spectroscopy method we presented.

\subsection{Photon shot noise spectra} 
\label{secTCLphoton}

We now apply the theory described in Appendix~\ref{secTCLideal} to the cQED experimental setting considered in the main text (see Fig.~\ref{figMicrograph}). In this setting, S comprises a pair of qubits encoded by the two lowest energy levels of transmons. The two qubits are coupled to a central bath consisting of a microwave resonator mode, which is itself subject to damping due to an external environment. We manipulate the qubits and the central bath by irradiating the input port of the common resonator with microwave drives. A continuous microwave drive with strength $\veps$ is first applied at frequency $\w_{\mrm d}$ near the fundamental resonator mode frequency $\omega_\mrm c$, in order to bring the mode into a coherent steady state. An additional pair of drives is then employed to initialize the qubits and measure their final state using the dispersive readout. Most importantly for this work, between initialization and readout, continuous spin-locking drives are also applied on each qubit $j$ with constant strength $\W_j$ at frequency $\w_{\mrm d j}$ near the bare qubit frequency $\w^0_{\mrm q j}$.

To describe the above experiment theoretically, we consider the sample parameters given in Table~\ref{tabSystemParams}, enabling us to make several simplifying assumptions.  First,  we assume that the qubit-resonator detunings $\D_j\equiv \w^0_{\mrm q j}-\w_{\mrm c}$ are large compared with: (i) $\veps$ and $\W_j$, enabling us to neglect any cross-talk effect of the resonator drive on the qubits and of the qubit drives on the resonator; and (ii) the qubit-resonator coupling strengths $g_j$, such that the two-qubit dispersive Hamiltonian can be employed~\cite{blais2004cavity}. Second,  we also assume that the qubit-qubit detuning $\w_{\mrm q 1}-\w_{\mrm q 2}$ is much larger than the strength of any interaction between the qubits, for example mediated by virtual transitions with the resonator~\cite{majer2007coupling}. We then model the two qubits, the resonator mode, and its environment with the Hamiltonian
\begin{align}
 &H_\mrm{QCE}(t)=\sum_{j=1,2}\left[\frac{\w_{\mrm q j}}2\s^z_j+\W_j\cos(\w_{\mrm d j}t)\s^x_j+\chi_j a^\dag a\s^z_j\right]\notag\\
  &\qquad+\w_\mrm c a^\dag a+2\veps \cos(\w_\mrm d t)(a+a^\dag)+H_\mrm{CE}+H_\mrm E,	
  \label{eqnHlab}
\end{align}
where $\chi_j\equiv g_j^2/\D_j$ is the dispersive coupling strength between qubit $j$ and the resonator mode with annihilation and creation operators $a$ and $a^\dag$, and $\w_{\mrm q j}=\w^0_{\mrm q j}+\chi_j$ is the Lamb-shifted qubit frequency. In Eq.~\eq{eqnHlab}, $H_\mrm{CE}$ is the Hamiltonian that describes coupling of the resonator mode with an external environment with free Hamiltonian $H_\mrm E$. Typically, this environment is modeled by an ensemble of harmonic oscillators corresponding to the modes of the free electromagnetic field or phonons by taking $H_\mrm{E}=\sum_k\w_k b^\dag_k b_k$ and $H_\mrm{CE}=\sum_k v_k a^\dag b_k+\mrm{H.c.}$, where $\w_k$ and $v_k$ are the frequency and coupling strength of the environmental mode $k$, whose excitations are annihilated by the bosonic operator $b_k$~\cite{walls2008quantum}.

We next move to the frame that rotates at the drive frequencies using the unitary tranformation 
\begin{align}
    R_\mrm d(t)=\exp\Big[-i\w_\mrm d t\,a^\dag a-i\sum_j\frac{\w_{\mrm dj}t}{2}\s^z_j\Big].
\end{align}
This transformation produces terms oscillating at frequencies $2\w_\mrm d$ and $2\w_{\mrm d j}$, which we neglect under the RWA, assuming that $\veps\ll\w_\mrm d$ and $\W_j\ll\w_{\mrm dj}$. In the rotating frame, the total Hamiltonian for the qubits, resonator, and environment is then given by Eq.~\eq{eqnHSCE}, in which
\begin{align}
    H_\mrm S&\rightarrow H'_\mrm S=\frac12\sum_j\left(\D'_{\mrm q j}\s^z_j+\W_j \s^x_j\right), \label{eqnHScQED}\\
    H_\mrm{SB}&\rightarrow H'_\mrm{SB}=\sum_j B_j\s^z_j,    \label{eqnHSBcQED}\\
    H_\mrm B&\rightarrow H'_\mrm B(t)=\D_{\mrm c} a^\dag a+\veps(a+a^\dag)+ H'_\mrm{CE}(t)+H_\mrm E,	\label{eqnHBcQED}
\end{align}
and where we have taken
\begin{align}
    \D'_{\mrm qj}\equiv \D_{\mrm q j}+2\chi_j\overline n,\hspace{1cm} B_j\equiv\chi_j (a^\dag a-\overline{n}).	\label{eqnBjcQED}
\end{align}
In Eq.~\eq{eqnHBcQED}, $\D_{\mrm q j}\equiv\w_{\mrm q j}-\w_{\mrm d j}$ and $\D_\mrm c\equiv \w_\mrm c-\w_\mrm d$ are the detunings of the qubit and resonator drives, respectively, and $H'_\mrm{CE}(t)\equiv R^\dag_\mrm d(t)H_\mrm{CE}R_\mrm d(t)=\sum_k v_ka^\dag b_k\eul{i\w_\mrm d t}+\mrm{H.c.}$ is the interaction-picture coupling between the resonator and its external environment. In addition, to arrive at a noise operator $B_j$ that has zero mean in the initial state, we have added and subtracted the term $\sum_j\chi_j\overline n\s^z_j$ in $H'_\mrm S$ and $H'_\mrm{SB}$, respectively, where $\overline{n}\equiv\mean{a^\dag a(0)}$ is the average photon number in the initial state of the resonator. In this approach, the mean of the noise is captured by the Stark shift $2\chi_j\overline n$ of each qubit $j$ in Eq.~\eq{eqnBjcQED} for $\D'_{\mrm q j}$. Assuming that this shift is measured with sufficient accuracy, we set $\D_{\mrm q j}=-2\chi_j\overline{n}$, leading to $\D'_{\mrm qj}=0\,\forall\,j$ in Eq.~\eq{eqnHScQED}. In the spin-locking basis, Eqs.~\eq{eqnHScQED} and \eq{eqnHSBcQED} then take the form of Eq.~\eq{eqnHSBspinLocking} we have taken in the derivations of Appendix~\ref{secTCLideal}.

We now follow the steps laid down in Appendix~\ref{secTCLideal} to obtain the reduced ME for the qubits. In Eq.~\eq{eqnHtilde}, we consider the interaction-picture bath operator $\tilde B_j(t)=R_\mrm B^\dag(t)B_jR_\mrm B(t)$, where $R_\mrm B(t)=\mathcal T\exp[-i\int_0^t ds\,H'_\mrm B(s)]$ is the evolution operator under $H'_\mrm B(t)$ given in Eq.~\eq{eqnHBcQED}, with $\mathcal T$ denoting time ordering. For weak coupling between the resonator mode and an environment consisting of a large number of degrees of freedom, the Born-Markov approximation may be invoked, following which reduced evolution of any central bath (resonator) operator under $R_\mrm B(t)$ is approximated by solving the Lindblad ME~\cite{carmichael1991open,walls2008quantum}, 
\begin{align}
 \dot\rho_\mrm C(t)&=\mathcal L_\mrm C\rho_\mrm C(t) \label{eqnMEresonator}\\
 \mathcal L_\mrm C \rho&\equiv-i[\D_\mrm c a^\dag a+\veps(a+a^\dag),\rho]+\kp\mathcal D[a]\rho. 	\label{eqnLresonator}
\end{align}
Assuming that the resonator couples to a continuum of environmental modes, the resonator damping rate appearing in Eq.~\eq{eqnLresonator} is given by $\kp = 2\pi D(\w_\mrm c)|v(\w_\mrm c)|^2$, where $D(\w)$ and $v(\w)$ are the frequency-dependent density of modes and coupling strength of the external environment, respectively. We also assume that the resonator drive is applied since time $t_0\ll-1/\kp$ so that, at $t=0$, the resonator mode has reached a steady state defined by $\dot \rho_\mrm C(t)=0$, in which noise is stationary. In this limit, Eq.~\eq{eqnMEresonator} is easily solved to give $\overline n=\mean{a^\dag a(0)}=\veps^2/[(\kp/2)^2+\D_\mrm c^2]$. 

Substituting Eq.~\eq{eqnMEresonator} into Eq.~\eq{eqnMultitime}, we may evaluate the central bath correlation functions $C_{jk}(t,s)$. In practice, to evaluate these correlators, we find that it is easier to use the equivalent quantum regression theorem~\cite{gardiner2000quantum}. This results in~\cite{blais2004cavity,clerk2010introduction}
\begin{align}
    C_{jk}(t)&=\chi_j\chi_k(\mean{a^\dag a(t)a^\dag a(0)}-\overline n^2)\notag\\
        &=\chi_j\chi_k\overline{n}\eul{-\kp|t|/2-i\D_{\mrm c}t}.
\end{align}
Fourier-transforming this correlation function then yields the shot-noise spectra given by Eq.~\eq{eqnSpectrumShotNoise}.

\subsection{Rabi-frequency difference and qubit relaxation} \label{secTCLrabiT1}

In Sec.~\ref{secSpinLocking} and Appendix~\ref{secTCLideal}, we considered an ideal setting in which noise only couples to $\s^z_j$, $j\in\{1,2\}$. In addition, we took the Rabi frequencies for the two spin-locking drives to be exactly equal, $\W_1=\W_2\equiv\W$. Here, we derive Eq.~\eq{eqnModifiedReduced}, which accounts for both a finite Rabi frequency difference ($\W_1\neq\W_2$) and noise coupling to the qubits off-axis, via $\s^x_j$, $j\in\{1,2\}$. Throughout this Appendix, we will assume that the Rabi frequencies $\W_1$ and $\W_2$ are approximately time-independent within any measurement of decay curves for a given target Rabi frequency $\W$, even though $\W_1$ and $\W_2$ are allowed to undergo small random fluctuations between distinct sets of measurements at different target values of $\W$.

To derive the ME, we follow the same steps as in Appendix~\ref{secTCLideal}, but replace $H_\mrm{SB}$ by $H^x_\mrm{SB}(t)\equiv\sum_j[B_j Q_j+B^x_j Q^x_j(t)]$ in the definition of the system-bath Hamiltonian, below Eq.~\eq{eqnHSCE}. In contrast with $H_\mrm{SB}$, $H^x_\mrm{SB}$ includes an additional central-bath operator $B^x_j$ that couples to the system through $Q^x_j(t)$. To describe $T_1$-effects, we take $Q^x_j(t)$ to be $\s^x_j$ in the frame co-rotating with the qubit drives, $Q^x_j(t)\equiv\s^+_j\eul{i\w_{\mrm d j}t}+\mrm{H.c.}$ Moving to the interaction picture with respect to $H_\mrm S+H_\mrm B$ as above, Eq.~\eq{eqnHtilde} is then replaced by
\begin{align}
 \tilde H^x_\mrm{SB}(t) = \sum_j\left[\tilde B_j(t)\tilde Q_j(t)+\tilde B^x_j(t)\tilde Q^x_j(t)\right].	\label{eqnHSBx}
\end{align}
Taking $H_\mrm S\rightarrow H'_\mrm S$, with $H'_\mrm{S}$ given by Eq.~\eq{eqnHSBspinLocking}, $\tilde Q^x_j(t)\equiv \eul{iH_\mrm S t}Q^x_j(t)\eul{-iH_\mrm S t}$ becomes, in the spin-locking basis,
\begin{align}
 \tilde Q^x_j(t)&=\frac12\left[\eul{i(\W_j+\w_{\mrm q j})t}-\eul{i(\W_j-\w_{\mrm q j}) t}\right]\tau^+_j+\mrm{H.c.}\notag\\
    &\qquad+\cos(\w_{\mrm q j} t)\tau^z_j,
\end{align}
where we have taken $\w_{\mrm d j}=\w_{\mrm q j}$, $\forall j$. We assume that $\mean{\tilde B^x_j(t)}=\Tr_\mrm B[\tilde B^x_j(t)\rho_\mrm B(0)]=0,$ $\forall t$, and that all cross-correlations involving any $\tilde B^x_j(t)$ vanish: $\mean{\tilde B^x_1(t_1)\tilde B^x_2(t_2)}=\mean{\tilde B^x_j(t_1)\tilde B_k(t_2)}=0$ $\forall j,k,t_1,t_2$. We also assume that noise due to $\tilde B^x_j(t)$ is stationary, and derive a TCL ME for $\tilde H^x_\mrm{SB}(t)$ similar to Eqs.~\eq{eqnMEkernel}-\eq{eqnKernel}, but in which several additional terms due to $\tilde B^x_j(t)$ and $\tilde Q^x_j(t)$ arise. Among these terms, some oscillate with angular frequencies $\W_j$ and $\w_{\mrm q j}\pm\W_j$. Since we must have $\W_j\ll\w_{\mrm q j},$ $\forall j$ (which we assumed in Sec.~\ref{secSpinLocking} to arrive at Eq.~\eq{eqnHSprime} within the RWA), we may neglect these fast oscillating terms within a secular approximation, as done below Eq.~\eq{eqnKernel}, without further loss of generality compared with Eq.~\eq{eqnReduced}. In addition, analogously to Eq.~\eq{eqnKernelSecular}, in the additional terms due to noise along $x$, noise spectra $S_{xx,j}(\w)\equiv\int_{-\infty}^\infty d\tau\,\eul{-i\w \tau}\mean{\tilde B^x_j(\tau)\tilde B^x_j(0)}$ are probed by filter functions in passbands of width $\sim 1/t$ about frequencies $\pm\w_{\mrm q j}$, $\w_{\mrm q j}\pm\W_j$, and $-\w_{\mrm q j}\pm\W_j$. Assuming that the noise spectra vary negligibly within these passbands, we take the infinite-time limit as in Appendix~\ref{secTCLideal}, and move back to the spin-locking reference frame rotating at the qubit drive frequencies. We then arrive at the ME 
\begin{align}
 &\dot\rho(t)=-i\sum_j\frac{\W_j}2\left[\tau^z_j,\rho(t)\right]+\sum_{jk}\left(\mathcal L^-_{jk}+\mathcal L^+_{jk}\right)\rho(t)\notag\\
  & +\sum_j\left(\frac{\g^z_j}{2}\mathcal D\left[\tau^z_j\right] + \g^-_j\mathcal D\left[\tau^-_j\right]+\g^+_j\mathcal D\left[\tau^+_j\right]\right)\rho(t), \label{eqnModifiedMEintermediate}
\end{align}
where $\mathcal L^\pm_{jk}$ are introduced in Eq.~\eq{eqnLjkplusminus} and
\begin{align}
    \g^z_j&\equiv \frac12\left[S_{xx,j}(\w_{\mrm q j})+S_{xx,j}(-\w_{\mrm q j})\right],   \label{eqngammaz}\\
    \g^\pm_j&\equiv\frac14\left[S_{xx,j}(\w_{\mrm q j}\pm\W_j) + S_{xx,j}(-\w_{\mrm q j}\pm\W_j)\right].	\label{eqngammaplusminus}
\end{align}

Allowing for $\W_1$ and $\W_2$ to be distinct, we introduce the Rabi frequency difference $\dt\W\equiv \W_1-\W_2$ and the average Rabi frequency $\W\equiv (\W_1+\W_2)/2$, so that $\W_1=\W+\dt\W/2$ and $\W_2=\W-\dt\W/2$. We then assume that $S_{jk}(\w)$ varies negligibly around $\w=\W$ over the range of values taken by $\dt\W$, allowing us to approximate $S_{jk}(\pm\W_1)\approx S_{jk}(\pm\W_2) \approx S_{jk}(\pm\W)$, $\forall j,k$. For the shot-noise spectra produced experimentally, this amounts to assuming that $\dt\W$ is negligible in comparison with the width $\kp$ of the peak, $\dt\W\ll\kp$. This enables us to approximate $\mathcal{L}^-_{jk}+\mathcal{L}^+_{jk}\approx\mathcal L_{jk}$ in Eq.~\eq{eqnModifiedMEintermediate}, with $\mathcal L_{jk}$ given by Eq.~\eq{eqnCorrelatedDecay}. Likewise, as in Ref.~\cite{yan2013rotating}, we assume that $S_{xx,j}(\w)$ varies negligibly over the small (typically, $\sim$~1-100~MHz) frequency ranges between $\pm\w_{\mrm q j}-\W_j$ and $\pm\w_{\mrm q j}+\W_j$ for all $j$, and thus take $S_{xx,j}(\w_{\mrm q j}\pm\W_j)\approx S_{xx,j}(\w_{\mrm q j})$ and $S_{xx,j}(-\w_{\mrm q j}\pm\W_j)\approx S_{xx,j}(-\w_{\mrm q j})$ in Eq.~\eq{eqngammaplusminus}. Under these approximations, Eq.~\eq{eqnModifiedMEintermediate} then reduces to Eq.~\eq{eqnModifiedReduced}, in which the damping rates due to noise along $x$ are given by
\begin{align}
    \G_{1,j}\equiv S_{xx,j}(\w_{\mrm q j})+S_{xx,j}(-\w_{\mrm q j})=1/T_1^{(j)},
\end{align}
with $T_1^{(j)}$ the longitudinal relaxation time of qubit $j$ in a free-evolution experiment.
The effects of noise along $x$ are then accounted for {\em phenomenologically} in the non-linear regression approach, by using the average $T_1^{(j)}$ values measured in independent free-evolution experiments.

Though a simultaneous characterization of all noise spectra $S_{jk}(\w)$ and $S_{xx,j}(\w)$ for all qubit axes would be much preferable in principle~\cite{paz-silva2019extending}, the above approach, in which the effect of $S_{xx,j}(\w)$ is captured by a single parameter $T_1^{(j)}$, has been successfully employed in a single-qubit context~\cite{yan2018distinguishing}. For sufficiently strong photon shot noise, longitudinal qubit relaxation effects are a significant, but weak correction that need only be taken into account at Rabi frequencies for which $S_{jk}(\w)$ is the weakest, i.e., at the tails of the Lorentzian spectra. Though not accounting for $T_1^{(j)}$ would make the reconstructed $S_{jk}(\w)$ to be significantly off at the tails, we do not expect the details of the additional noise spectra $S_{xx,j}(\w)$ to lead to significant effects beyond corrections obtained with Eq.~\eq{eqnModifiedReduced}.

\section{Confidence intervals for spectrum estimates}	
\label{secConfidenceIntervals}

M-estimators were introduced in the 1960s for robust estimation of a parameter using data whose distribution function is only approximately known: for example, the observations may follow a normal distribution, except for a fraction of them which is affected by experimental error~\cite{huber1964robust}. Following the steps of Ref.~\cite{van2000asymptotic}, in this Appendix we discuss the asymptotic normality of these estimators when the number of observations tends to infinity. This will lead us to the approximate confidence intervals for the spectrum values presented in the main text. No excessive emphasis is put on mathematical rigor; the following steps are valid under suitable regularity conditions that an interested reader may find in Ref.~\cite{van2000asymptotic}.

Given a loss function $\ld(z)$ that penalizes deviation between a model and experimental observations, the general form of an M-estimator $\hat{\mvec\q}$ for a $p$-dimensional parameter vector $\mvec \q$ is
\begin{align}
 \hat{\mvec\q}=\argmin_{\mvec \q}\sum_{\al=1}^d \ld(z_\al),	\label{eqnDefM}
\end{align}
where $z_\al$ is the $\alpha$-th realization of a random variable $Z$ associated with experimental observations, and whose probability distribution is parameterized by $\mvec \q$. In Sec.~\ref{secEstimation}, the parameter $\mvec\q$ is the spectrum vector, $\mvec\q=\mvec S$, while in Sec.~\ref{secValidation}, $\mvec \q$ is the spectrum vector supplemented by the Rabi frequency difference $\dt\W$ between the qubit drives. Throughout the main text, $z_\al$ quantifies the relative deviation of observations from their expected value $\mean{O_\al}_{\mvec \q}$ via $z_\al=(\overline O_\al-\mean{O_\al}_{\mvec \q})/\overline \s_\al$, where $\overline O_\al$ is the sample mean of $M$ projective measurements performed on the two-qubit system and $\overline\sigma_\al^2=\mrm{var}(\overline O_\al)$. Throughout this Appendix, we will assume that all realizations $z_\al$ are independent and identically distributed (i.i.d.). For example, under ideal conditions (in the absence of any experimental error), and in the asymptotic limit in which each $\overline O_\al$ is obtained from $M\to\infty$ projective measurements, the central limit theorem implies that each $z_\al$ is sampled from the standard normal distribution $\mathcal N(0,1)$.

Since the value of $\mvec\q$ that minimizes the right-hand side of Eq.~\eq{eqnDefM} may be found by setting derivatives with respect to $\q_\ell$ to zero, M-estimators are often alternatively defined by the solution of a set of $p$ \emph{estimating equations}
\begin{align}
 \sum_{\al=1}^d\y_\ell(z_\al)=0,\hspace{5mm}\ell\in\{1, 2, \ldots, p\}.
\end{align}
To describe the minimization problem of Eq.~\eq{eqnDefM}, above, we set $\y_\ell(z_\al)\equiv\partial \ld(z_\al)/\partial\q_\ell$. It is also convenient to introduce the $d$-dimensional column vectors $\mvec\y(z_\al)$ and $\mvec\Y(\mvec\q)$, whose components $\ell$ are defined as $\y_\ell(z_\al)$ and $\Y_\ell(\mvec\q)\equiv\sum_{\al=1}^d \y_\ell(z_\al)$, respectively. With this notation, the estimating equations for the M-estimator of $\mvec\q$ may be written succinctly as
\begin{align}
 \mvec\Y(\hat{\mvec \q})=0.	\label{eqnEstimatingEquationsSuccinct}
\end{align}
To discuss the asymptotic behavior of $\hat{\mvec \q}$, we Taylor-expand Eq.~\eq{eqnEstimatingEquationsSuccinct} about $\hat{\mvec \q}=\mvec \q^\ast$, where $\mvec\q^\ast$ is the true value of $\mvec\q$, and truncate to the first order. This gives
\begin{align}
 \mvec\Y(\hat{\mvec\q})\approx\mvec\Y(\mvec\q^\ast)+\dot{\mvec\Y}(\mvec\q^\ast)(\hat{\mvec\q}-\mvec\q^\ast),	\label{eqnTaylor}
\end{align}
where $\dot{\mvec\Y}(\mvec\q^\ast)$ is the $p\times p$ matrix of derivatives of $\mvec\Y(\mvec \q)$ at $\mvec\q=\mvec\q^\ast$, namely, 
\begin{align}
 \dot{\Y}_{k\ell}(\mvec\q^\ast)\equiv\left.\del{\Y_k(\mvec\q)}{\q_\ell}\right|_{\mvec\q=\mvec\q^\ast}=\sum_{\al=1}^d\left.\del{\y_k(z_\al)}{\q_\ell}\right|_{\mvec\q=\mvec\q^\ast}.	\label{eqnDefPsiDot}
\end{align}
By definition of M-estimators, Eq.~\eq{eqnEstimatingEquationsSuccinct}, the left-hand side of Eq.~\eq{eqnTaylor} must vanish, $\mvec\Y(\hat{\mvec \q})=0$. Assuming that $\dot{\mvec\Y}(\mvec\q^\ast)$ is invertible then gives
\begin{align}
 \hat{\mvec\q}-\mvec\q^\ast\approx-\dot{\mvec\Y}(\mvec\q^\ast)^{-1}\mvec\Y(\mvec\q^\ast).	\label{eqnTaylorTheta}
\end{align}
Recall that $\dot{\mvec\Y}(\mvec\q^\ast)$ and $\mvec\Y(\mvec\q^\ast)$ are ultimately functions of the realizations $z_\al$ of the random variable describing the deviation of an experimental observation $\overline O_\al$ from its expected value. We thus investigate the asymptotic behavior of $\dot{\mvec\Y}(\mvec\q^\ast)$ and $\mvec\Y(\mvec\q^\ast)$ as the number of observations tend to infinity, $d\to\infty$, assuming that the $z_\al$'s are i.i.d. Since, by definition, $\Y_\ell(\mvec\q^\ast)=\sum_{\al=1}^d\y_\ell(z_\al)$, the central limit theorem implies the following convergence in distribution:
\begin{align}
 \mvec\Y(\mvec\q^\ast)=\sum_{\al=1}^d\mvec\y(z_\al)\to\sqrt d\,\mathcal{N}_p(0,\mrm{cov}(\mvec\y)),	\label{eqnAsyPsi}
\end{align}
where $\mathcal N_p(\mvec\mu,\mvec\Sg)$ is the $p$-dimensional multivariate normal distribution with mean $\mvec \mu$ and covariance matrix $\mvec \Sg$. To arrive at Eq.~\eq{eqnAsyPsi}, we have used the fact that $\mathbb E[\mvec\y]=0$ at $\mvec\q=\mvec\q^\ast$, where $\mathbb E(\mvec \y)$ is the expectation value of $\y$ over realizations of the random variable $Z$. This property must be satisfied for any M-estimator since, by the weak law of large numbers, $\mathbb E[\mvec\y]=\lim_{d\to\infty}\frac1d\sum_{\al=1}^d\mvec\y(z_\al)$, where $\sum_{\al=1}^d\mvec\y(z_\al)=0$ at $\mvec\q=\hat{\mvec\q}$ by Eq.~\eq{eqnEstimatingEquationsSuccinct}, and $\lim_{d\to\infty} \hat{\mvec \q}=\mvec\q^\ast$ due to asymptotic consistency of M-estimators~\cite{van2000asymptotic}. Because $\mathbb E[\mvec\y]=0$, the covariance matrix in Eq.~\eq{eqnAsyPsi} is simply given by $\mrm{cov}(\mvec\y)=\mathbb E(\mvec\y\mvec\y^T)$.

In addition, applying the weak law of large numbers to the last expression in Eq.~\eq{eqnDefPsiDot} leads to the following convergence in probability:
\begin{align}
 \dot{\mvec\Y}(\mvec\q^\ast)\to d\,\mathbb E(\dot{\mvec \y}),	\label{eqnAsyPsiDot}
\end{align}
where we have introduced the matrix $\dot{\mvec \y}(z_\al)$, whose components are $\dot{\y}_{k \ell}(z_\al)\equiv\left.\partial\y_k(z_\al)/\partial\q_\ell\right|_{\mvec\q=\mvec\q^\ast}$. 

\begin{figure*}[t]
\includegraphics[width=0.95\textwidth]{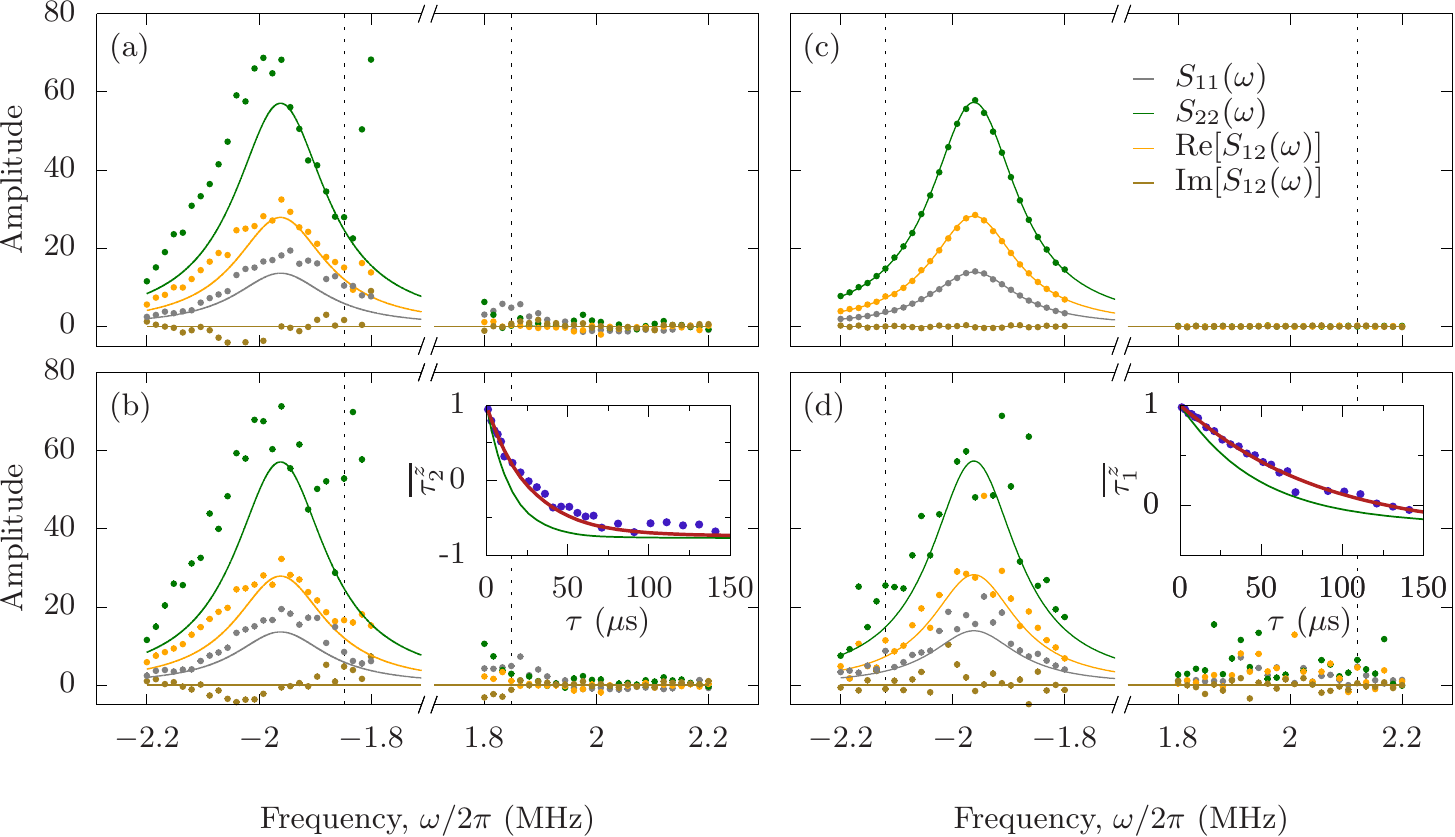}
 \caption{{\em Robust estimation approach.}
 Comparison of spectral reconstructions using Huber loss (top row) and weighted least squares (bottom row). {\bf (a)} Reconstructions of the spectra from experimental data using the Huber loss function. {\bf (b)} Same reconstructions using weighted least squares. {\bf (c)} Reconstructions from simulated data using the Huber loss function. {\bf (d)} Same reconstructions using weighted least squares. Solid lines: ideal spectra given by Eq.~\eq{eqnSpectrumShotNoise}. Gray, green, yellow and brown curves correspond to $S_{11}(\w)$, $S_{22}(\w)$, $\mrm{Re}[S_{12}(\w)]$, and $\mrm{Im} [S_{12}(\w)]$, respectively. Inset of (b): decay curve for the sample mean $\overline{\tau^z_2}$ as a function of evolution time with $\W/2\pi=1.848$~MHz. Inset of (d): decay curve for simulated sample means of $\overline{\tau^z_1}$ with $\W/2\pi=2.120$~MHz. Blue dots: experimental data [inset of (a)] or simulated data [inset of (d)]. Solid lines: non-linear regression with Huber loss (red) and weighted least squares (green). In both insets, the initial qubit state is $\ket{{+x},{+x}}$. Dashed lines in the main plots indicate spectrum frequencies $\w/2\pi=\pm\W/2\pi$ corresponding to the insets.
 \label{fig:Huber}}
 \end{figure*}
 
 Substituting Eqs.~\eq{eqnAsyPsi} and \eq{eqnAsyPsiDot} into Eq.~\eq{eqnTaylorTheta}, applying Slutsky's theorem \cite{casella2002statistical}, and using the linear transformation property $\mrm{cov}(\mvec B\mvec x)=\mvec B\mrm{cov}(\mvec x)\mvec B^T$, where $\mvec x$ is a random column vector and $\mvec B$ a constant matrix, then implies that $\hat {\mvec\q}-\mvec \q^\ast$ converges in distribution to
\begin{align}
 \hat{\mvec \q}-\mvec\q^\ast\to\mathcal{N}_p\left(0,\mvec\Sg^{\mvec\q}\right),
\end{align}
with the covariance matrix
\begin{align}
 \mvec\Sg^{\mvec\q}=\frac1d\,\mathbb E\left(\dot{\mvec \y}\right)^{-1}\mathbb E\left(\mvec\y\mvec\y^T\right)\left[\mathbb E\left(\dot{\mvec\y}\right)^{-1}\right]^T,	\label{eqnCovarianceMestimator}
\end{align}
where $\mvec\q$ is evaluated at $\mvec \q^\ast$. 

We may now use the definition $\y_\ell(z_\al)\equiv\partial\ld(z_\al)/\partial\q_\ell$ to approximate the covariance matrix of $\hat{\mvec\q}-\mvec\q^\ast$ in terms of derivatives of loss functions for $d\gg1$. Evaluating the derivatives using the chain rule and approximating expectation values of functions of observations $f(Z)$ by $\mathbb E(f)\approx\frac1d\sum_{\al=1}^d f(z_\al)$ for $M\gg1$ then gives
\begin{align}
 \mathbb E[\mvec\y\mvec\y^T]&\approx \frac1d\mvec J^T\mvec D^2\mvec J,	\label{eqnpsipsiT}\\
 \mathbb E[\dot{\mvec\y}]_{k \ell}&\approx \frac1d(\mvec J^T\mvec \D\mvec J)_{k\ell}+\frac1d\sum_\al\left.D_{\al\al}\frac{\partial^2z_\al}{\partial \q_k\partial\q_\ell}\right|_{\mvec\q=\mvec\q^\ast},	\label{eqnpsidotkl}
\end{align}
where we have introduced the Jacobian matrix with elements $J_{\al \ell}\equiv \left.\partial z_\al/\partial \q_\ell\right|_{\mvec \q=\mvec\q^\ast}$, and where $\mvec \D$ and $\mvec D$ are diagonal matrices whose non-zero elements are $D_{\al\al}=\left.\partial\ld/\partial z\right|_{z=z_\al}$ and $\D_{\al\al}=\left.\partial^2\ld/\partial z^2\right|_{z=z_\al}$, respectively.
These derivatives are readily obtained for simple loss functions. In particular, for quadratic loss, $\ld(z)=z^2/2$, we have $D_{\al\al}=z_\al$ and $\D_{\al\al}=1$, whereas for Huber loss, the following hold:
\begin{align}
 D_{\al\al}=\left\{\begin{array}{ll}
                    z_\al,	&|z_\al|\leq\dt_0,\\
                    \dt_0\mrm{sign}(z_\al),	&\mbox{otherwise,}
                   \end{array}
 \right.
 \hspace{1mm}
 \D_{\al\al}=\left\{\begin{array}{ll}
                    1,	&|z_\al|\leq\dt_0,\\
                    0,	&\mbox{otherwise.}
                   \end{array}
 \right.
\end{align}

Substituting Eq.~\eq{eqnpsipsiT} and \eq{eqnpsidotkl} into Eq.~\eq{eqnCovarianceMestimator} gives an approximation of the covariance matrix $\mvec\Sg^{\mvec\q}$ of $\hat{\mvec\q}-\mvec\q^\ast$ for $d\gg1$. We then estimate $95\%$ confidence intervals presented in the main text from
\begin{align*}
 \hat\q_\ell\pm1.96\sqrt{\Sg^{\mvec \q}_{\ell\ell}}.
\end{align*}

 For a sufficiently weakly non-linear relationship between $z_\al$ and the parameters, the term proportional to second-order derivatives in Eq.~\eq{eqnpsidotkl} may be neglected, leading to the compact expression
\begin{align}
 \mvec \Sg^{\mvec\q}\approx(\mvec J^T\mvec \D\mvec J)^{-1}(\mvec J^T \mvec D^2\mvec J)(\mvec J^T\mvec \D\mvec J)^{-1}.	\label{eqnSgWeakNonLinearity}
\end{align}
Though we do account for the term containing second-order derivatives in all confidence-interval calculations presented in the main text for completeness, in all cases studied here, this term is found to be negligible. Since it avoids the computation of second-order derivatives of $z_\al$ with respect to all possible pairs of $\q_k$ and $\q_\ell$, evaluation of the confidence intervals through Eq.~\eq{eqnSgWeakNonLinearity} is much more efficient numerically, leading to significant savings in computation time.

\section{Comparison of reconstructions with weighted least squares vs. Huber loss functions} 
\label{secHuber}

Spectral estimation based on weighted least squares is particularly vulnerable to outliers in experimental data, thus motivating robust estimation strategies such as M-estimation. In this Appendix, we further illustrate the adverse effect of outliers by comparing two-qubit spectrum reconstructions obtained with the robust Huber loss function with estimates based on weighted least squares, considering both experimental and simulated data. 

Figure \ref{fig:Huber} displays estimates of the two-qubit spectra obtained from Eq.~\eq{eqnSest} using the Huber loss function (top row) with the tuning parameter $\dt_0=1$, along with estimates obtained using weighted least squares (bottom row). In Fig.~\ref{fig:Huber}(a) and Fig.~\ref{fig:Huber}(b), we show reconstructions using the experimental data discussed in Sec.~\ref{secValidation}. A close inspection of these figures reveals that reconstructions obtained by using weighted least squares are significantly more noisy, an effect that is particularly visible in $S_{22}(\omega)$ between $\omega/2\pi=-2.0$~MHz and $\omega/2\pi=-1.8$~MHz. To verify that this effect stems from the fitting procedure, we plot an example of decay curve -- here, $\overline{\tau^z_2}$ -- as a function of evolution time for $\W/2\pi= 1.848$~MHz as an inset in Fig.~\ref{fig:Huber}(b). This inset reveals that while the fit using Huber loss closely follows the bulk of the data points, the fit using weighted least squares goes astray, most plausibly due to the influence of outliers in the many decay curves involved in the global regression procedure. As indicated by the dashed vertical lines in Fig.~\ref{fig:Huber}(a)-(b), the reconstructed spectra at the corresponding frequencies $\w/2\pi=\pm1.848$~MHz lie closer to the theoretical value using Huber loss than with weighted least squares, in particular for $S_{22}(\w)$. 

To verify that such a failure of weighted least squares can indeed arise from outliers, we reproduce the effect with simulated data in Fig.~\ref{fig:Huber}(c)-(d). To produce the simulated data used in the spectral estimation, we first calculate the two-qubit density matrix $\rho(t)$ by substituting the spectra given by Eq.~\eq{eqnSpectrumShotNoise} into the ME defined by Eqs.~\eq{eqnReduced}-\eq{eqnCorrelatedDecay}, which we solve numerically for the shot-noise parameters measured in the experiment (caption of Fig.~\ref{figSpectrum}). From the resulting $\rho(t)$, we evaluate probabilities associated with binary outcomes ($\pm1$) of projective measurements, and generate $M=2000$ simulated outcomes for each observable $O_r$ by sampling a Bernoulli distribution. We then evaluate the sample means $\overline O_\al$ of projective measurement outcomes for each combination of initial state, observable, and evolution time given in Table~\ref{tabExample}, and their corresponding standard deviations. Finally, to emulate experimental error, we use a $\dt$-contaminated model~\cite{casella2002statistical}, in which each value of $\overline O_\al$ is assigned a probability 0.1 to be replaced by an outlier, which we model by sampling a uniform probability distribution between $-1$ and $+1$.

After constructing this simulated data set, we perform the reconstructions using the procedure explained in Sec.~\ref{secEstimation}. Despite the presence of numerous outliers, Fig.~\ref{fig:Huber}(c) shows that our procedure successfully reconstructs all the spectra when using Huber loss. However, in Fig.~\ref{fig:Huber}(d), the equivalent reconstructions using weighted least squares are remarkably more noisy, and even involve a spurious positive-frequency component. The discrepancy between outcomes of the two loss functions is significantly more pronounced than in the experiment, as may be expected for the rather pessimistic model of outliers employed here. The inset of Fig.~\ref{fig:Huber}(d) displays the outcome of the fitting procedure for a pair of frequencies at which weighted least squares failed to produce accurate spectrum estimates [$\W/2\pi=2.120$~MHz, dashed vertical lines in Fig.~\ref{fig:Huber}(c)-(d)]. Again, while the decay curve fitted using weighted least squares wanders away, Huber loss produces a decay curve that closely follows the bulk of the data.

\bibliography{spinLocking.bib}

\end{document}